\documentclass[10pt,journal,twocolumn]{IEEEtran}
\usepackage{amsmath,amsfonts}
\usepackage{amsthm}
\usepackage{amssymb}
\usepackage{easyReview}
\usepackage{algorithmic}
\usepackage{algorithm}
\usepackage{array}
\usepackage[caption=false,font=normalsize,labelfont=sf,textfont=sf]{subfig}
\usepackage{textcomp}
\usepackage{stfloats}
\usepackage{url}
\usepackage{bm}
\usepackage{verbatim}
\usepackage{graphicx}
\usepackage{cite}
\usepackage{amsthm}
\usepackage{caption}
\usepackage{booktabs}
\makeatletter
\renewcommand{\maketag@@@}[1]{\hbox{\m@th\normalsize\normalfont#1}}%
\makeatother

\allowdisplaybreaks[4]
\newtheorem{Theorem}{Theorem}
\newtheorem*{Proof}{Proof}

\begin{document}

\title{Joint Association, Beamforming, and Resource Allocation for Multi-IRS Enabled MU-MISO Systems With RSMA
}

\author{Chunjie Wang, Xuhui Zhang, Huijun Xing, Liang Xue,~\IEEEmembership{Member,~IEEE,} Shuqiang Wang,~\IEEEmembership{Senior Member,~IEEE,} Yanyan Shen,~\IEEEmembership{Member,~IEEE,} Bo Yang,~\IEEEmembership{Senior Member,~IEEE,} and Xinping Guan,~\IEEEmembership{Fellow,~IEEE}
\thanks{C. Wang is with the Shenzhen Institute of Advanced Technology, Chinese Academy of Sciences, Shenzhen 518055, China, the School of Information and Electrical Engineering, Hebei University of Engineering, Handan 056038, China, and also with the Hebei Key Laboratory of Security and Protection Information Sensing and Processing, Handan 056038, China (e-mail: cj.wang@siat.ac.cn)}
\thanks{X. Zhang is with the Shenzhen Future Network of Intelligence Institute, the School of Science and Engineering, and the Guangdong Provincial Key Laboratory of Future Networks of Intelligence, The Chinese University of Hong Kong, Shenzhen, Guangdong 518172, China (e-mail: xu.hui.zhang@foxmail.com).
}
\thanks{H. Xing is with Department of Electrical and Electronic Engineering, Imperial College London, London SW7 2AZ, The United Kingdom, and is also with the Shenzhen Future Network of Intelligence Institute, the Guangdong Provincial Key Laboratory of Future Networks of Intelligence, The Chinese University of Hong Kong, Shenzhen, Guangdong 518172, China (e-mail: huijunxing@link.cuhk.edu.cn).
}
\thanks{L. Xue is with the School of Information and Electrical Engineering, Hebei University of Engineering, Handan 056038, China (e-mail: lxue17@asu.edu)}
\thanks{S. Wang and Y. Shen are with the Shenzhen Institute of Advanced Technology, Chinese Academy of Sciences, Shenzhen 518055, China (e-mail: sq.wang@siat.ac.cn; yy.shen@siat.ac.cn)}
\thanks{B. Yang and X. Guan are with the Department of Automation and the Key Laboratory of System Control and Information Processing, Ministry of Education, Shanghai Jiao Tong University, Shanghai 200240, China (e-mail: bo.yang@sjtu.edu.cn; xpguan@sjtu.edu.cn).}
\thanks{
Notice: This work has been submitted to the IEEE for possible publication. Copyright may be transferred without notice, after which this version may no longer be accessible.
}
}

\maketitle

\begin{abstract}
Intelligent reflecting surface (IRS) and rate-splitting multiple access (RSMA) technologies are at the forefront of enhancing spectrum and energy efficiency in the next generation multi-antenna communication systems. This paper explores a RSMA system with multiple IRSs, and proposes two purpose-driven scheduling schemes, i.e., the exhaustive IRS-aided (EIA) and opportunistic IRS-aided (OIA) schemes. The aim is to optimize the system weighted energy efficiency (EE) under the above two schemes, respectively. Specifically, the Dinkelbach, branch and bound, successive convex approximation, and the semidefinite relaxation methods are exploited within the alternating optimization framework to obtain effective solutions to the considered problems. The numerical findings indicate that the EIA scheme exhibits better performance compared to the OIA scheme in diverse scenarios when considering the weighted EE, and the proposed algorithm demonstrates superior performance in comparison to the baseline algorithms.
\end{abstract}


\section{Introduction}
\IEEEPARstart {T}{he} embracing of the era of artificial intelligence (AI), especially the AI generated content, and large language models, brings the large-scale data communication requirement in emerging applications, such as autonomous driving, augmented reality, virtual reality, image automatic processing, and industrial AI integration, towards the next generation communication system.
To meet the demand of operations of emerging technologies,
and improve the diverse quality of service of connected mobile users, the beyond fifth-generation / sixth-generation (B5G/6G) cellular networks should support significantly increased spectrum efficiency, ultra-low transmission
latency, and extremely-high communication reliability \cite{8808168}.
There are currently several methods currently being utilized to improving the capacity of communication systems, one of which is multi-antenna technology. By designing and coordinating transmit/receive beamforming, spectrum utilization efficiency can be greatly improved \cite{9246254}.
Hence, the studies of integrated multiple antenna technology with wireless systems have been widely adopted in \cite{9644606, 8537962,9918632,9154573}.

 The burgeoning requirement for higher data rates in communication systems has given rise to innovative multiple access technologies, such as rate-splitting multiple access (RSMA), to driven the development of the system throughput and efficiency. RSMA, with its transmitter-based linear or non-linear rate splitting and receiver employing successive interference cancellation (SIC), offers a flexible and resilient approach for designing and optimizing non-orthogonal transmission, multi-access, and interference management strategies in future multi-antenna wireless systems. Through its user-data splitting mechanism, RSMA enables the simultaneous transmission of distinct users' data within a single time slot in the same spectrum, resulting in higher data transmission rates and spectral efficiencies \cite{artirsma,als20216}. Beyond enhancing the data transmission rate, the RSMA technique also amplifies the signal-to-noise ratio. This enhancement significantly bolsters the reliability and robustness of the communication systems \cite{9831440}. Evidently, RSMA is an comprehensive and versatile technique that subsumes or generalizes four seemingly distinct techniques: orthogonal multiple access, non-orthogonal multiple access (NOMA), space division multiple access (SDMA), and physical-layer multicasting. It is noteworthy that while RSMA simplifies these strategies under certain conditions, it generally outperforms them. RSMA's innovation has made it a groundbreaking research field, and it is replete with immense potential for future high-efficiency communication systems \cite{9832611,9226406,9663192}.\\
\indent Intelligent reflecting surface (IRS)
is an innovative technology that is expected to revolutionize the future of B5G/6G communication networks.
The core function of IRS is to improve wireless link performance through software programming, which adjusts both the reflection phase and amplitude of the incoming signal based on feedback from communication link information, thus improving the signal coverage, enhancing the desired signal strength, mitigating the undesired interference signals and noises from environment, reshaping the rank of desired communication channel, and promoting the communication capacity of the worst-case users (e.g., the users located at the network edge) \cite{9519632,10440056}.
Additionally,
by deploying IRS in wireless networks, a new degree of spatial freedom can be achieved, paving the way for the realization of intelligent wireless programmable environments \cite{8910627,9907933,9326394}.
Due to the aforementioned attributes, the IRS can be deemed as a large-scale passive antenna array, surpassing traditional wireless communication equipment in terms of power efficiency and reliability, such as amplify-and-forward relay \cite{9095301}. Therefore, the IRS has been studied extensively and incorporated into various wireless systems \cite{9623452,9751048,10288203,9417469,10497119}.\\
\IEEEpubidadjcol
\indent Furthermore, the compelling benefits of IRS and RSMA have generated significant interest in integrating IRS into RSMA systems, leading to an emerging research area. \cite{9832618} provides a thorough examination of the integration of RSMA with IRS technologies. The study demonstrates that the IRS-enabled RSMA framework achieves a superior rate region and spectral efficiency compared to both IRS-assisted SDMA and NOMA schemes. \cite{9849099} explored a MISO system optimized for minimizing transmit power consumption through the integration of IRS and RSMA. In \cite{10000454}, an RSMA architecture assisted by an aerial IRS in a downlink MISO scenario with imperfect SIC was investigated. The transmit beamforming and user-common rate optimization were formulated to enhance the total achievable rate. And an IRS-enabled unmanned aerial vehicle-based RSMA system was examined in \cite{9760044}.

As we all know, one of the most prominent virtues of IRS is its low power consumption. Nevertheless, when multiple IRSs comprising a significant number of reflecting elements coexist within the system, the energy consumption of the IRSs cannot be neglected. Therefore, given the presence of non-negligible power consumption of multiple IRSs, investigating the system energy efficiency (EE) proves to be a valuable direction. The enhancement of EE in IRS-enabled RSMA systems has been explored in \cite{10008582,10032267,9145189}. Specifically, \cite{10008582} investigated the short-packet communications within an IRS-enabled RSMA system, with the objective of optimizing system EE through the jointly design of the base station precoder and IRS's passive beamforming. \cite{10032267} discussed the issue of maximizing EE while considering the transmitter's power constraints, user rate requirements, and energy harvesting constraints. The study jointly optimized the transmit beamforming, common user rates, power splitting ratios, and IRS's phase shifts. Similarly, in \cite{9145189}, enhancing system EE was achieved by adjusting the IRS phase shifts and the transmit beamforming.

However, to date, there have been only a few studies focusing on RSMA systems assisted by multiple IRSs. Hence, we present a multi-IRS assisted multi-user multiple-input single-output (MU-MISO) system RSMA system, and we also 
propose two scheduling mechanisms for the IRSs, namely the exhaustive IRS-aided (EIA) and opportunistic IRS-aided (OIA) schemes. More precisely, in the former approach, all IRSs within the system are scheduled to enhance communication between the access point (AP) and any user. Whereas in the latter, the IRSs and the users are associated opportunistically. Subsequently, we investigate the weighted EE maximization problem of the system in both schemes, respectively. We address the non-convex optimization challenge within the EIA scheme using methods such as Dinkelbach, semidefinite relaxation (SDR), and successive convex approximation (SCA), supported by an alternating optimization (AO) framework. These methods jointly optimize the common rate for the users, the AP's transmit beamforming, and the phase shift matrices of the IRSs. For the OIA scheme, we employ a suite of methods—including Dinkelbach, branch and bound (BnB), SCA, and SDR—within the AO framework to solve the original problem iteratively. Here, we optimize the association coefficient between the IRSs and users, the AP's transmit beamforming, the phase shift matrices of the IRSs, and the common rate for the users.

The main contributions of this work are summarized as follows:
\begin{itemize}
	\item Firstly, multiple IRSs are introduced into the RSMA system, along with the proposal of two distinct schemes for scheduling these IRSs in the system, namely EIA and OIA schemes. The weighted EE maximization problem is explored under both the EIA and OIA schemes.  It is noteworthy that the problem under the EIA scheme is non-convex, while the problem under the OIA scheme is a mixed integer nonlinear programming (MINLP) problem.
	\item Secondly,
 we apply various mathematically creative tactics to address the optimization problems introduced by the EIA and OIA scheduling schemes.
 Tactics such as the Dinkelbach method, branch and bound algorithm, successive convex approximation, and semidefinite relaxation methods are utilized to transform the original problems into tractable forms. Solutions to these tractable form problems are then obtained iteratively, enhancing the efficiency and effectiveness of the optimization process.
	\item Finally, simulation outcomes confirm the superiority of the EIA scheme over the OIA scheme in terms of EE performance. The proposed algorithms are shown to outperform the baseline algorithms, demonstrating their efficacy in optimizing the utilization of multiple IRSs in the RSMA system. These results not only validate the proposed scheduling schemes but also underline the significance of the optimization techniques employed in improving system EE.
\end{itemize}

\textit{Organizations:}
The rest of this paper is organized as follows.
Section II introduces the multi-IRS assisted
MU-MISO system with RSMA protocol. 
Section III formulates the weighted EE maximization problem for the EIA scheme and proposes an efficient solution.
Section IV introduces the weighted EE maximization problem for the OIA scheme and designs an efficient solution.
Section V presents numerical results to validate the performance of our proposed EIA scheme and OIA scheme  compared to other benchmark schemes. 
Finally, Section VI concludes this paper.
Besides, two proofs of the corresponding theorems are listed in the Appendix A and Appendix B.

\textit{Notations:}
The notation used in this paper is elaborated below. $ {{\mathbb{C}}^{M\times N}} $ represents the $ M \times N $ complex matrix $ \mathbb{C} $. $ {\cal C}{\cal N}(\mu ,{\sigma ^2}) $ denotes the circularly symmetric complex Gaussian distribution with a mean of $ \mu $ and variance $ {\sigma ^2} $. The symbol $ \mathrm{j} $ signifies the imaginary unit, such that $ {\mathrm{j}^2} = -1 $. For a generic matrix $ {\bf{G}} $, $ {{\bf{G}}^{\mathrm{H}}} $, $ {{\bf{G}}^{\mathrm{T}}} $, and $ \rm{Tr}({\bf{G}}) $ stand for the conjugate transpose, the transpose, and the trace of $ {\bf{G}} $, respectively. In the case of a vector $ {\bf{w}} $, $ ||{\bf{w}}|| $ signifies the Euclidean norm. The notation $ \rm{diag}({\bf{w}}) $ indicates a diagonal array where the elements along the diagonal correspond to those in vector $ {\bf{w}} $. The optimal value of an optimization variable $ x $ is represented by $ x^\star $.

\section{System Model}
\begin{figure}[t]
	\centering
	\includegraphics[width=0.95\linewidth]{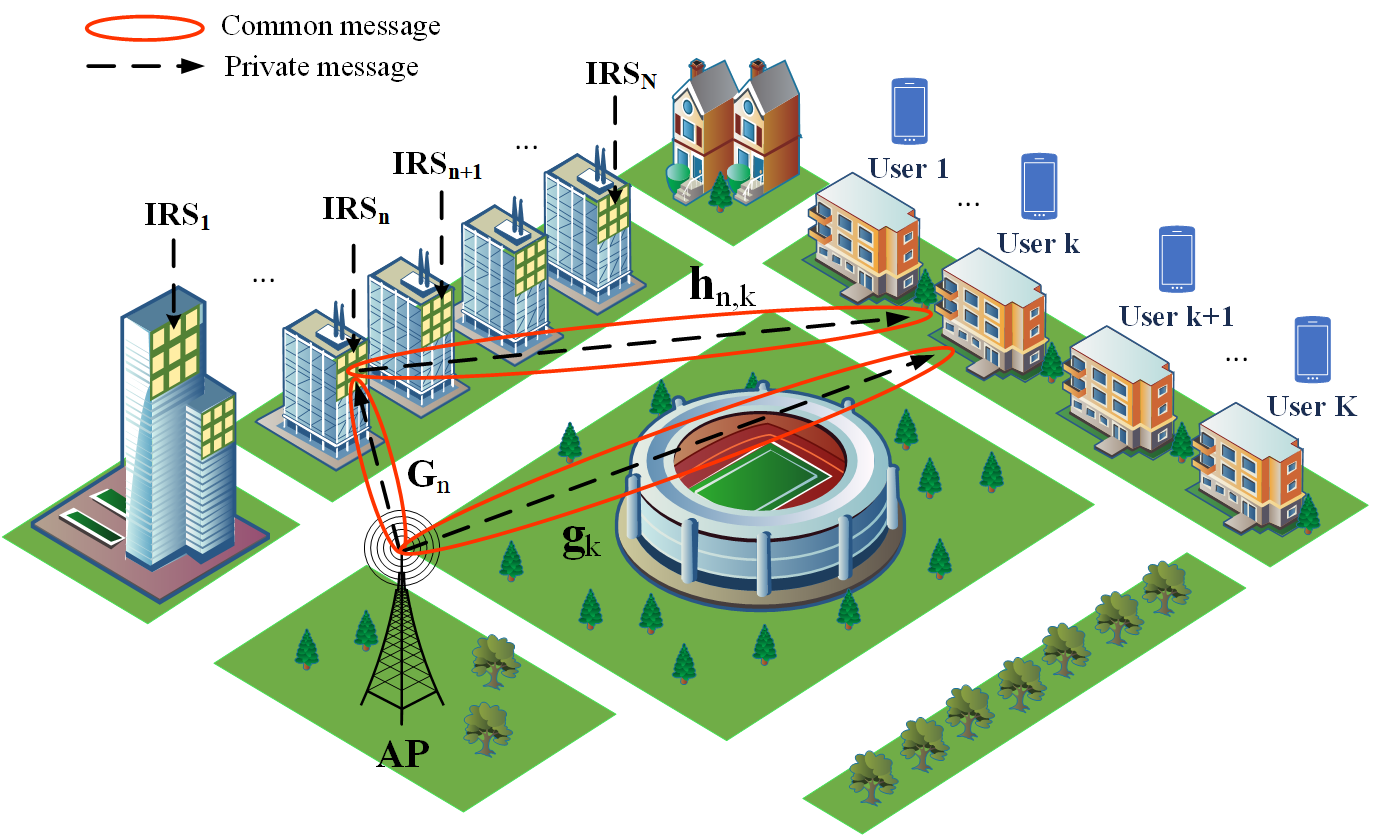}
	\captionsetup{justification=centering}
	\caption{The RSMA system model assisted by multiple IRSs.}
	\label{Fig:1}
\end{figure}

As depicted in Fig. \ref{Fig:1},
we investigate
a multi-IRS assisted MU-MISO system with RSMA. The architecture of this system comprises a single AP equipped with $ M $ antennas, $ K $ users each equipped with single antenna, and $ N $ IRSs, each comprising $ L $ reflecting elements. The deployment of IRSs aims to augment communication in light of obstacles such as buildings and trees obstructing direct channels between the AP and users. In this paper, we explore a composite link that integrates direct and cooperative links, assuming perfect knowledge of the channel state information \cite{8811733,10000454}.
\subsection{IRS Model}

$ {{\bf{\Theta}} _n} = \sqrt \eta {\rm{diag}}\left(\theta _n^1,\theta _n^2,...,\theta _n^L\right) \in {\mathbb{C}^{L \times L}} $ is denoted as the phase shift matrix of the $ n $-th IRS, where $ \theta _n^l = {e^{j\varphi _n^l}} $, $ \varphi _n^l $ in which represents the phase shift of $ l $-th reflecting element in the $ n $-th IRS, and $ \varphi _n^l \in [0,2\pi ) $. The reflection efficiency $ \eta $ is set to $ 1 $ for simplicity.

\subsection{Signal Model}
\subsubsection{The Exhaustive IRS-aided Scheme}
Within this scheme, all the IRSs are scheduled to assist the intercommunication between the AP and users. According to the one-layer rate-splitting scheme, we can divide the message intended for the $ k $-th user into a common message and a private message \cite{9832618, 9831440}. A shared codebook encodes the common messages into a common stream $ s_c $, while each user $ k $ has a dedicated stream $ s_k $ for their private message. Hence, the transmitted signal at the AP takes the form
\begin{gather}
	{\bf{x}} = {\bf{w}}_c{s_c}+\sum\limits_{k = 1}^K {{{\bf{w}}_k}{s_k}},
\end{gather}
where $ {\bf{w}}_c \in {\mathbb{C}^{M \times 1}} $ denotes the transmit beamforming vector for the common message, and $ {\bf{w}}_k \in {\mathbb{C}^{M \times 1}} $ represents the corresponding private beamforming vector for user $ k $. In EIA scheme, the signal received at user $ k $ is 
\begin{gather}
	y_k^{\mathrm{EIA}} = \big({\bf{g}}_k^{\mathrm{H}} + \sum\limits_{n = 1}^N {{\bf{h}}_{n,k}^{\mathrm{H}}{{\bf{\Theta}} _n}} {{\bf{G}}_n}\big){\kern 1pt} {\kern 1pt} {\bf{x}} + n_k^{\mathrm{EIA}},
\end{gather}
where $ {\bf{g}}_k \in {\mathbb{C}^{M \times 1}}$ is the channel vector between the AP and user $ k $, $ {{{\bf{h}}}_{n,k}} \in {\mathbb{C}^{L \times 1}}$ represents the channel vector between the IRS $ n $ and user $ k $, $ {\bf{G}}_n \in {\mathbb{C}^{L \times M}}$ denotes the channel matrix from the AP to IRS $ n $. The additive white Gaussian noise (AWGN) is denoted as $ {n_k^{\mathrm{EIA}}} \sim \mathcal{C}\mathcal{N}(0,\sigma _{k}^{2}) $.

The SIC technique is utilized in RSMA systems to separate common and private messages. Initially, the common message is extracted by treating all the private messages as noise. Following the decoding of the common message, the private message of a specific user can be recovered by regarding the other private messages as interference. Let ${\bf{H}}_k^{\mathrm{H}} = {\bf{g}}_k^{\mathrm{H}} + \sum\limits_{n = 1}^N {{\bf{h}}_{n,k}^{\mathrm{H}}{{\bf{\Theta }}_n}{{\bf{G}}_n}}$, the achievable rates for decoding the common and private messages at user $ k $ are respectively expressed as
\begin{align}
	R_{c,k}^{\mathrm{EIA}} = {\log _2}\big( {1 + \frac{{{{\left| {{\mathbf{H}}_k^{\mathrm{H}}{{\mathbf{w}}_c}} \right|}^2}}}{{\sum\limits_{i = 1}^K {{{\left| {{\mathbf{H}}_k^{\mathrm{H}}{{\mathbf{w}}_i}} \right|}^2}}  + \sigma _k^2}}} \big),\label{eq:3}
\end{align}
and
\begin{align}
	R_{p,k}^{\mathrm{EIA}} = {\log _2}\big( {1 + \frac{{{{\left| {{\mathbf{H}}_k^{\mathrm{H}}{{\mathbf{w}}_k}} \right|}^2}}}{{\sum\limits_{i = 1,i \ne k}^K {{{\left| {{\mathbf{H}}_k^{\mathrm{H}}{{\mathbf{w}}_i}} \right|}^2}}  + \sigma _k^2}}} \big),\label{eq:4}
\end{align}

\noindent To ensure all users can successfully decode $ s_c $, the system common rate can not exceed each common rate decoded by each user, that is
\begin{align}
	R_c^{\mathrm{EIA}} = \min \left\{ {R_{c,k}^{\mathrm{EIA}},\forall k} \right\}.
\end{align}

\noindent Denote $ C_k^{\mathrm{EIA}} $ as the allocated segment of the common rate $ R_c^{\mathrm{EIA}} $ designated for user $ k $, we can derive the following constraint
\begin{align}
	\sum\limits_{k = 1}^K {{C_k^{\mathrm{EIA}}} } \le R_c^{\mathrm{EIA}},
\end{align}

\noindent Therefore, the achievable rate for user $ k $ within the EIA scheme is
\begin{align}
	R_k^{\mathrm{EIA}} = {C_k^{\mathrm{EIA}}} + R_{p,k}^{\mathrm{EIA}}.
\end{align}

\subsubsection{The Opportunistic IRS-aided Scheme}
To reduce the resource usage, we propose the OIA scheme. In this scheme, we denote $ {{\mathbf{v}}_{n,k}} = \left\{ {{\varpi _{n,k}},\forall n,k} \right\} $ the association matrix containing the association coefficient $ {\varpi _{n,k}} $. The coefficient variable $ {\varpi_{n,k}}=1 $ if user $ k $ is linked to IRS $ n $, and $ {\varpi_{n,k}} = 0 $ otherwise. To promote the adoption of multi-IRS setups, we impose a constraint on the maximum number of users that can be connected to the same IRS, and then obtain the following constraint
\begin{align}
	\sum\limits_{k = 1}^K {{\varpi _{n,k}}}  \le a, \forall n,
\end{align}
where $ a $ is a constant. 

Undoubtedly, the presence of multiple IRSs provides the users with a variety of choices. However, in the OIA scheme, only a single IRS can be chosen to help the communication between the AP and any specific user, which implies
\begin{align}
	\sum\limits_{n = 1}^N {{\varpi _{n,k}}}  = 1, \forall k,
\end{align}

\noindent The signal received by user $ k $ with the aid of IRS $ n $ in the OIA scheme can be expressed as 
\begin{gather}
	y_{n,k}^{{\mathrm{{OIA}}}} = \left({\bf{g}}_k^{\mathrm{H}} + {\bf{h}}_{n,k}^{\mathrm{H}}{{\bf{\Theta}} _n}{{\bf{G}}_n}\right) {\bf{x}} + n_k^{{\mathrm{{OIA}}}},
\end{gather}
where $ {n_k^{\mathrm{{OIA}}}}\sim \mathcal{C}\mathcal{N}(0,\delta _{k}^{2}) $ represents the AWGN. Let $ \widetilde {\bf{H}}_{n,k}^{\mathrm{H}} = {\bf{g}}_k^{\mathrm{H}} + {\bf{h}}_{n,k}^{\mathrm{H}}{{\bf{\Theta}} _n}{{\bf{G}}_n} $, the achievable rates for user $ k $ in decoding the common and private messages, aided by IRS $ n $, are respectively expressed as
\begin{align}
	R_{c,n,k}^{\mathrm{OIA}} = {\log _2}\big( {1 + \frac{{{{\left| {\widetilde {\mathbf{H}}_{n,k}^{\mathrm{H}}{{\mathbf{w}}_c}} \right|}^2}}}{{\sum\limits_{i = 1}^K {{{\left| {\widetilde {\mathbf{H}}_{n,k}^{\mathrm{H}}{{\mathbf{w}}_i}} \right|}^2}} + \delta _k^2}}} \big),
\end{align}
and
\begin{align}
	R_{p,n,k}^{\mathrm{OIA}} = {\log _2}\big( {1 + \frac{{{{\left| {\widetilde {\mathbf{H}}_{n,k}^{\mathrm{H}}{{\mathbf{w}}_k}} \right|}^2}}}{{\sum\limits_{i = 1,i \ne k}^K {{{\left| {\widetilde {\mathbf{H}}_{n,k}^{\mathrm{H}}{{\mathbf{w}}_i}} \right|}^2}} + \delta _k^2}}} \big).
\end{align}

\noindent The common message decoding rate is
\begin{align}
	R_c^{\mathrm{OIA}} = \min \left\{ {{\varpi _{n,k}} R_{c,n,k}^{\mathrm{OIA}} \big\vert {\varpi _{n,k}} = 1,\forall n,k } \right\}.
\end{align}

\noindent The constraint on $ C_k^{\mathrm{OIA}} $, the allocation of the common rate $ R_c^{\mathrm{OIA}} $ to user $ k $, is expressed as follows
\begin{align}
	\sum\limits_{k = 1}^K {C_k^{\mathrm{OIA}} } \le R_c^{\mathrm{OIA}}.
\end{align}

\noindent Ultimately, the total achievable rate of user $ k $ with the assistance of IRS $ n $ in the OIA scheme is given by
\begin{align}
	R_{n,k}^{\mathrm{OIA}} = C_k^{\mathrm{OIA}} + R_{p,n,k}^{\mathrm{OIA}}.
\end{align}

\subsection{Power Consumption Model}
This paper examines a power consumption model similar to that in \cite{huang2019reconfigurable}. The system's overall power usage comprises the AP's transmit power and the static power drawn by the AP, users, and IRSs.

Under the EIA scheme, the system's total power consumption is
\begin{align}
	{\cal P}_{\mathrm{total}}^{{\mathrm{EIA}}} &= {\left\| {{{\bf{w}}_c}} \right\|^2} + \sum\limits_{k = 1}^K {{{\left\| {{{\bf{w}}_k}} \right\|}^2}} + \sum\limits_{k = 1}^K {{P_{\mathrm{user},k}}}  + {P_{AP}} \notag \\ &+ N L {P}(b),
\end{align}
where $ P_{AP} $ and $ P_{user,k} $ denote the static power usage by the AP and user $ k $, respectively. $ P(b) $ represents the power consumption of each reflecting element at the IRSs with $ b $-bit resolution.

In the OIA scheme, the comprehensive system power consumption amounts to
\begin{align}
	{\cal P}_{\mathrm{total}}^{{\mathrm{OIA}}} &= {\left\| {{{\bf{w}}_c}} \right\|^2} + \sum\limits_{k = 1}^K {{{\left\| {{{\bf{w}}_k}} \right\|}^2}} + \sum\limits_{k = 1}^K {{P_{\mathrm{user},k}}}  + {P_{\mathrm{AP}}} \notag \\& + \sum\limits_{n = 1}^N {\max \{\varpi _{n,k},\forall k\}} L {P}(b).
\end{align}
where {\small $ \sum\limits_{n = 1}^N {\max \{\varpi _{n,k},\forall k\}} $} denotes the number of selected serving IRSs.
\subsection{EE Design}
Based on all the aforementioned assumptions, the advantage of the OIA over EIA scheme is that it reduces the users' signal processing overhead, because in the OIA scheme, there are fewer reflected signals from IRSs need to be processed at each user. Specifically, the number of signals from the IRSs and AP for each user is $ (NL+1) $ in the EIA scheme, but the number reduces to $ (L+1) $ in the OIA scheme. In addition, the power consumption of the IRSs in the OIA scheme is generally less than that in the EIA scheme. These variances create a notable disparity in EE performance between the two schemes.

The EE represents the ratio of the overall achievable rate to the total system power consumption. In addition, to ensure the fairness among users, we study the weighted EE in this paper. Then, the weighted EE expressions in the EIA and OIA schemes can be respectively represented as
\begin{gather}
	{\eta _{\mathrm{EE}}^{\mathrm{EIA}}} = \frac{B\sum\limits_{k = 1}^K {{\upsilon _k}}R_k^{\mathrm{EIA}}}{{\cal P}_{\mathrm{total}}^{{\mathrm{EIA}}}},
\end{gather}
and
\begin{gather}
	{\eta _{\mathrm{EE}}^{\mathrm{OIA}}} = \frac{B\sum\limits_{n = 1}^N\sum\limits_{k = 1}^K {{\upsilon _k}}{\varpi _{n,k}}R_k^{\mathrm{OIA}}} {{{\cal P}_{\mathrm{total}}^{{\mathrm{OIA}}}}},
\end{gather}
where $ B $ signifies the system bandwidth, $ {\upsilon _k} \ge 0 $ serves as the weighted coefficient of user $ k $, and a greater value of $ {\upsilon _k} $ signifies a significantly higher priority for user $ k $.

\section {Problem Formulation and Solution for the EIA Scheme}\label{section:3}
In this section, we optimize the AP transmit beamforming $ {{\bf{w}}_c}, \{ {{\bf{w}}_k},\forall k \} $, the phase shift matrices $ \{ {\bf{\Theta}}_n,\forall n \} $ at the IRSs, and the common rate allocation coefficients $ \{ C^{\mathrm{EIA}}_k, \forall k \} $ to maximize weighted EE for the EIA scheme.

Denote $ {\bf{w}} = [{{\bf{w}}_c},{{\bf{w}}_1},{{\bf{w}}_2},...,{{\bf{w}}_K}] $, {\small $ {{\bm{\theta}} _n} = {[\theta _n^1,\theta _n^2,...,\theta _n^L]^T} $}, {\small $ {\bm{\theta}} = [{{\bm{\theta}} _1},{{\bm{\theta}} _2},...,{{\bm{\theta}} _N}] $} and {\small $ {{\bf{C}}^{\mathrm{EIA}}} = [C_1^{\mathrm{EIA}},...,C_K^{\mathrm{EIA}}] $}, the optimization problem can be expressed mathematically as
\begin{subequations}
	{\small \begin{align}
		{\rm{P1}}:\ \mathop {\max }\limits_{{\bf{w}},{\bm{\theta}},{{\mathbf{C}}^{\mathrm{EIA}}}} \quad &\frac{{B\sum\limits_{k = 1}^K {{\upsilon _k}{R_k^{\mathrm{EIA}}}} }}{{{\cal P}_{\mathrm{total}}^{{\mathrm{EIA}}}}}, \label{eq:20a} \\
		\mathrm{s.t.} \qquad\ &{\left\| {{{\mathbf{w}}_c}} \right\|^2} + \sum\limits_{k = 1}^K {{{\left\| {{{\bf{w}}_k}} \right\|}^2}} \le {P_{\max }},\label{eq:20b} \\
		&R_k^{\mathrm{EIA}} \ge {R_{k }},\forall k,\label{eq:20c}\\
		&\sum\limits_{k = 1}^K {{C_k^{\mathrm{EIA}}} \le } R_{c,k}^{\mathrm{EIA}}, \forall k, \label{eq:20d} \\
		&C_k^{\mathrm{EIA}} \ge 0,\forall k,\label{eq:20e} \\
		&\left| {\theta _n^l} \right| = 1, \forall n,\ l, \label{eq:20f}
	\end{align}}%
\end{subequations}
where $ R_{k} $ is the user $ k $'s minimum required rate, $ P_{\max} $ is the AP's maximum transmit power. Constraints (\ref{eq:20b}), (\ref{eq:20c}), (\ref{eq:20d}), and (\ref{eq:20f}) correspond to power constraints at the AP, minimum rate requirements for users, total common rate control, and phase shift constraints at the IRSs, respectively.

Problem P1 is characterized as non-convex, arising from its non-convex fractional objective function, non-convex constraint (\ref{eq:20f}), and the complex coupling relationship between the IRS phase shifts and the transmit beamforming vector ${\bf{w}}$ within constraints (\ref{eq:20c}) and (\ref{eq:20d}). Solution is obtained by initially reframing the problem into a more tractable format, which is then divided into two subproblems using an AO-based approach. Solutions to these subproblems are obtained in an iterative manner.

\subsection{Problem Transformation}
The fractional objective function (\ref{eq:20a}) can be transformed into a subtractive form through the application of the Dinkelbach method \cite{9133120,9882159}, which is reformulated as
{\small \begin{align}
		F(\rho_1 ) &= B\sum\limits_{k = 1}^K {{\upsilon _k}} \big(C_k^{\mathrm{EIA}}+{\log _2}( {1 + \frac{{{{\left| {{\mathbf{H}}_k^{\mathrm{H}}{{\mathbf{w}}_k}} \right|}^2}}}{{\sum\limits_{i = 1,i \ne k}^K {{{\left| {{\mathbf{H}}_k^{\mathrm{H}}{{\mathbf{w}}_i}} \right|}^2}} + \sigma _k^2}}} )\big)\notag
		\\& - \rho_1 \mathcal{P}_{\mathrm{total}}^{\mathrm{EIA}},\label{eq:21}
\end{align}}%
where $ \rho_1 $ is an auxiliary variable. 
Then, by substituting (\ref{eq:21}) into problem P1, the optimization problem can be equivalently transformed into a more convenient form
\begin{subequations}
		\begin{align}
			{\rm{P2}}:&\mathop {\max }\limits_{{\mathbf{w}},{\bm{\theta }}, {{\mathbf{C}}^{\mathrm{EIA}}}}\ F(\rho_1), \\
			&\quad\ \mathrm{s.t.}\quad\ \ {\text{(\ref{eq:20b})} - \text{(\ref{eq:20f})}} .
		\end{align}
\end{subequations}

\noindent By denoting {\small $ {{\mathbf{A}}_k} = \big[ {{\text{diag}}({\mathbf{h}}_{1,k}^{\mathrm{H}}){{\mathbf{G}}_1};...;{\text{diag}}({\mathbf{h}}_{N,k}^{\mathrm{H}}){{\mathbf{G}}_N};{\mathbf{g}}_k^{\mathrm{H}}} \big] $ and $ {\mathbf{f}} = [{{\mathbf{f}}_1};...;{{\mathbf{f}}_n};...;{{\mathbf{f}}_N};1] $}, where
$ {\mathbf{f}}_n^{\mathrm{H}} = [f_n^1,...,f_n^L] $ and $ f_n^l = {e^{\mathrm{j}\varphi _n^l}},\forall n,l $, we can obtain
\begin{align}
	{\mathbf{H}}_k^{\mathrm{H}}{{\mathbf{w}}_\tau } = {{\mathbf{f}}^{\mathrm{H}}}{{\mathbf{A}}_k}{{\mathbf{w}}_\tau }, \tau  = c,1,...,K,\label{eq:23}
\end{align}

\noindent By substituting (\ref{eq:23}) into problem P2, the rewritten form of P2 is
{\begin{subequations}
		\begin{align}
			&{\rm{P3}}:\mathop {\max }\limits_{{\mathbf{w}},{\bf{f }}, {{\mathbf{C}}^{\mathrm{EIA}}}}\ {F^{'}}(\rho_1 ),\\
			&\mathrm{s.t.}\ C_k^{\mathrm{EIA}} + {\log _2}\big(1 + \frac{{{{\left| {{{\mathbf{f}}^{\mathrm{H}}}{{\mathbf{A}}_k}{{\mathbf{w}}_k}} \right|}^2}}}{{\sum\limits_{i = 1,i \ne k}^K {{{\left| {{{\mathbf{f}}^{\mathrm{H}}}{{\mathbf{A}}_k}{{\mathbf{w}}_i}} \right|}^2}} + \sigma _k^2}}\big) \ge {R_{k}},\forall k,\\
			&\qquad \sum\limits_{k = 1}^K {C_k^{\mathrm{EIA}}}  \le {\log _2}\big(1 + \frac{{{{\left| {{{\mathbf{f}}^{\mathrm{H}}}{{\mathbf{A}}_k}{{\mathbf{w}}_c}} \right|}^2}}}{{\sum\limits_{i = 1}^K {{{\left| {{{\mathbf{f}}^{\mathrm{H}}}{{\mathbf{A}}_k}{{\mathbf{w}}_i}} \right|}^2}}  + \sigma _k^2}}\big),\forall k,\\
			&\qquad \left| {f_n^l} \right| = 1,\forall n,l,\\
			&\qquad \text{(\ref{eq:20b}),(\ref{eq:20e})},
		\end{align}
\end{subequations}
where 
{\small \begin{align}
		{F^{'}}(\rho_1 ) &= B\sum\limits_{k = 1}^K {{\upsilon _k}} \big(C_k^{\mathrm{EIA}}+{\log _2}(1 + \frac{{{{\left| {{{\mathbf{f}}^{\mathrm{H}}}{{\mathbf{A}}_k}{{\mathbf{w}}_k}} \right|}^2}}}{{\sum\limits_{i = 1,i \ne k}^K {{{\left| {{{\mathbf{f}}^{\mathrm{H}}}{{\mathbf{A}}_k}{{\mathbf{w}}_i}} \right|}^2}} + \sigma _k^2}})\big)\notag
		\\& - \rho_1 \mathcal{P}_{\mathrm{total}}^{\mathrm{EIA}}.
\end{align}}%

\noindent Subsequently, utilizing the AO-based algorithm, we jointly optimize the transmit beamforming $ {\mathbf{w}} $ and the common rate allocation $ {{\mathbf{C}}^{\mathrm{EIA}}} $ for given $ {\bf{f}} $ at first. After that, we optimize the phase shift matrix $ {\bf{f}} $ with obtained $ {\mathbf{w}} $ and $ {{\mathbf{C}}^{\mathrm{EIA}}} $.

\subsection{Common Rate Allocation and Transmit Beamforming Optimization}

With the phase shift matrix ${\mathbf{f}}$ fixed, the common rate allocation and transmit beamforming $\{ {\mathbf{w}}, {\mathbf{C}}^{\mathrm{EIA}} \}$ are optimized jointly. Let {\small $ {{\mathbf{W}}_\tau } = {{\mathbf{w}}_\tau }{\mathbf{w}}_\tau ^{\mathrm{H}} $}, then it follows that {\small $ {{\mathbf{W}}_\tau } \succcurlyeq {\mathbf{0}} $} and {\small $ {\text{Rank}}({{\mathbf{W}}_\tau }) = 1, \tau=c,1,...,K $}. Denote {\small $ {{\mathbf{B}}_k} = {\mathbf{A}}_k^{\mathrm{H}}{\mathbf{f}}{{\mathbf{f}}^{\mathrm{H}}}{{\mathbf{A}}_k} $} and {\small $ {\mathbf{W}} = [{{\mathbf{W}}_c},{{\mathbf{W}}_1},{{\mathbf{W}}_2},...,{{\mathbf{W}}_K}] $}, then the optimization problem P3 with variables {\small $ \{{\mathbf{W}}, {\mathbf{C}}^{\mathrm{EIA}}\} $} can be represented as
\begin{subequations}
	{\small \begin{align}
			{\text{P4}}: \mathop {\max }\limits_{{\mathbf{W}},{{\mathbf{C}}^{\mathrm{EIA}}}} \ &B\sum\limits_{k = 1}^K {{\upsilon _k}\big(C_k^{\mathrm{EIA}} + {U_{p,k}}({\mathbf{W}}) + {V_{p,k}}({\mathbf{W}}) \big)}\notag \\
			& - \rho_1 \big({\text{Tr}}({{\mathbf{W}}_c}) + \sum\limits_{k = 1}^K {{\text{Tr}}({{\mathbf{W}}_k})} + {\Delta ^{\mathrm{EIA}}}\big),\label{eq:26a} \\
			\mathrm{s.t.}\quad\ &{\text{Tr}}({{\mathbf{W}}_c}) + \sum\limits_{k = 1}^K {{\text{Tr}}({{\mathbf{W}}_k})}  \le {P_{\max }}, \\
			&C_k^{\mathrm{EIA}} + {U_{p,k}}({\mathbf{W}}) + {V_{p,k}}({\mathbf{W}}) \ge {R_{k}},\forall k,\label{eq:26c} \\
			&\sum\limits_{k = 1}^K {C_k^{\mathrm{EIA}}}  \le {X_{c,k}}({\mathbf{W}}) + {Y_{c,k}}({\mathbf{W}}),\forall k,\label{eq:26d} \\
			&C_k^{\mathrm{EIA}} \ge 0,\forall k,\\
			&{{\mathbf{W}}_\tau } \succcurlyeq {\mathbf{0}},\tau  = c,1,...,K,\label{eq:26f} \\
			&{\text{Rank}}({{\mathbf{W}}_\tau }) = 1,\tau  = c,1,...,K,\label{eq:26g}
	\end{align}}%
\end{subequations}
where
{\small \begin{align}
		&{U_{p,k}}({\mathbf{W}}) = {\log _2}\big(\sum\limits_{i = 1}^K {{\text{Tr}}({{\mathbf{B}}_k}{{\mathbf{W}}_i})} + \sigma _k^2\big),\forall k,\qquad\qquad\quad \label{eq:27} \\
		&{V_{p,k}}({\mathbf{W}}) = - {\log _2}\big(\sum\limits_{i = 1,i \ne k}^K {{\text{Tr}}({{\mathbf{B}}_k}{{\mathbf{W}}_i})} + \sigma _k^2\big),\forall k,\label{eq:28}\\
		&{X_{c,k}}({\mathbf{W}}) = {\log _2}\big(\sum\limits_{i = 1}^K {{\text{Tr}}({{\mathbf{B}}_k}{{\mathbf{W}}_i}) + } {\text{Tr}}({{\mathbf{B}}_k}{{\mathbf{W}}_c}) + \sigma _k^2\big),\forall k,\label{eq:29}\\
		&{Y_{c,k}}({\mathbf{W}}) = - {\log _2}\big(\sum\limits_{i = 1}^K {{\text{Tr}}({{\mathbf{B}}_k}{{\mathbf{W}}_i})} + \sigma _k^2\big),\forall k,\label{eq:30}\\
		&{\Delta ^{\mathrm{EIA}}} = \sum\limits_{k = 1}^K {{P_{\mathrm{user},k}}} + {P_{\mathrm{AP}}} + N L {P}(b).
\end{align}}%

We can observe that constraints (\ref{eq:26c}) and (\ref{eq:26d}) as well as the objective function are non-convex due to the concave nature of (\ref{eq:27}) and (\ref{eq:29}) and the convexity of (\ref{eq:28}) and (\ref{eq:30}). This implies that the optimization problem P4 is non-convex. Noted that these constraints are represented as differences of concave functions, which can be approximated through the utilization of their corresponding lower-bound functions after the application of the following theorem\footnote{It is worth noting that the Taylor formula can also be employed to address this issue. However, both approaches display substantial differences in the solving process. Theorem \ref{Theorem:1} allows for a direct determination of the maximum value through mathematical analysis, obviating the need for approximation or derivative calculations, hence ensuring precise results.}.

\begin{Theorem}\label{Theorem:1}
	{\rm{\cite{9195771,7015632}}} : Let $ a \in {\mathbb{R}^{1 \times 1}} $ be a positive variable, $ b $ be a constant, and $ f(a) = - (ab/\ln 2) + {\log _2}a + (1/\ln 2) $, then we can obtain
\begin{align}
	- {\log _2}b = \mathop {\max }\limits_{a > 0} f(a),\label{eq:32}
\end{align}
The optimal solution to the expression on the right-hand side of {\rm{(\ref{eq:32})}} is $ a = 1/b $.
\end{Theorem} 
\begin{Proof}
    The theorem's validity can be demonstrated by setting $ \frac{{\partial f(a)}}{{\partial a}} $ to zero.
    $\hfill \square$
\end{Proof}

In the $ t $-th iteration, following Theorem \ref{Theorem:1}, around the point $ {\textbf{W}^{(t-1)}} $ obtained in the $ (t-1) $-th iteration, we substitute the expression for $ V_{p,k}(\textbf{W}) $ in (\ref{eq:26a}) and (\ref{eq:26c}) with its corresponding lower bound, which is given by
{\small \begin{align}
	V_{p,k}^{L,(t)}({{\mathbf{W}}^{(t)}})&= \mathop {\max }\limits_{\varrho _{p,k}^{(t)} > 0} - \frac{{\varrho_{p,k}^{(t)}\big(\sum\limits_{i = 1,i \ne k}^K {{\text{Tr}}({{\mathbf{B}}_k}{{\mathbf{W}}_i^{(t)}})} + \sigma _k^2\big)}}{{\ln 2}}\notag \\
	& + {\log _2}\varrho_{p,k}^{(t)} + (1/\ln 2),\label{eq:33}
\end{align}}%
where
{\small \begin{align}
	\varrho_{p,k}^{(t)} = {\big(\sum\limits_{i = 1,i \ne k}^K {{\text{Tr}}({{\mathbf{B}}_k}{\mathbf{W}}_i^{(t - 1)})} + \sigma _k^2\big)^{ - 1}}.\label{eq:34}
\end{align}}%

\noindent Similarly, denote {\small $ Y_{c,k}^{L,(t)}({{\mathbf{W}}^{(t)}}) $} as the global underestimation of {\small $ {Y_{c,k}}({\mathbf{W}}) $}, then we have
{\small \begin{align}
	Y_{c,k}^{L,(t)}({{\mathbf{W}}^{(t)}}) &= \mathop {\max }\limits_{\varrho_{c,k}^{(t)} > 0} - \frac{{\varrho_{c,k}^{(t)}\big(\sum\limits_{i = 1}^K {{\text{Tr}}({{\mathbf{B}}_k}{{\mathbf{W}}_i^{(t)}})} + \sigma _k^2\big)}}{{\ln 2}}\notag \\
	& + {\log _2}\varrho_{c,k}^{(t)} + (1/\ln 2),\label{eq:35}
\end{align}}%
where 
{\small \begin{align}
	\varrho_{c,k}^{(t)} = {\big(\sum\limits_{i = 1}^K {{\text{Tr}}({{\mathbf{B}}_k}{\mathbf{W}}_i^{(t - 1)})} + \sigma _k^2\big)^{ - 1}}.\label{eq:36}
\end{align}}

\noindent A lower bound solution to problem P4 is found by solving problem P5 in $ t $-th iteration
\begin{subequations}
	{\small \begin{align}
			&{\text{P5}}: \mathop {\max }\limits_{{{\mathbf{W}}^{(t)}},{{\mathbf{C}}^{\mathrm{EIA}}}} B\sum\limits_{k = 1}^K {\upsilon _k}\big( V_{p,k}^{L,(t)}({{\mathbf{W}}^{(t)}}) + C_k^{\mathrm{EIA}} + {U_{p,k}}({{\mathbf{W}}^{(t)}}) \big)\notag \\
			&\quad\ \ - \rho_1 \big({\text{Tr}}({\mathbf{W}}_c^{(t)}) + \sum\limits_{k = 1}^K {{\text{Tr}}({\mathbf{W}}_k^{(t)})} + {\Delta ^{\mathrm{EIA}}}\big), \\
			&\mathrm{s.t.}\ \ {\text{Tr}}({\mathbf{W}}_c^{(t)}) + \sum\limits_{k = 1}^K {{\text{Tr}}({\mathbf{W}}_k^{(t)})} \le {P_{\max }}, \label{eq:37b}\\
			&\qquad C_k^{\mathrm{EIA}} + {U_{p,k}}({{\mathbf{W}}^{(t)}}) + V_{p,k}^{L,(t)}({{\mathbf{W}}^{(t)}}) \ge {R_{k}},\forall k, \label{eq:37c}\\
			&\qquad \sum\limits_{k = 1}^K {C_k^{\mathrm{EIA}}} \le {X_{c,k}}({{\mathbf{W}}^{(t)}}) + Y_{c,k}^{L,(t)}({{\mathbf{W}}^{(t)}}),\forall k, \label{eq:37d}\\
			&\qquad C_k^{\mathrm{EIA}} \ge 0,\forall k,\\
			&\qquad {\mathbf{W}}_\tau ^{(t)} \succcurlyeq {\mathbf{0}},\tau  = c,1,...,K, \label{eq:37f}\\
			&\qquad {\text{Rank}}({\mathbf{W}}_\tau ^{(t)}) = 1,\tau  = c,1,...,K.\label{eq:37g}
	\end{align}}%
\end{subequations}

Observing the non-convexity of problem P5 lies in the rank-one constraint (\ref{eq:37g}), we employ the SDR method to tackle this issue. Subsequently, the tightness of this relaxation strategy is validated through a subsequent theorem.

\begin{Theorem}\label{Theorem:2}
	The solution to problem P5 is always satisfies {\small $ {\text{Rank}}({\mathbf{W}}_\tau) = 1, \tau = c,1,...,K $}. 
\begin{Proof}
	Please refer to the Appendix A.
 $\hfill \square$
\end{Proof}
\end{Theorem}

\subsection{Phase Shift Optimization}

Utilizing the obtained $ \bf{w} $ and $ {\bf{C}}^{\mathrm{EIA}} $, by canceling out irrelevant terms in P3, the optimization problem pertaining to $ \bf{f} $ can be articulated as follows
\begin{subequations}
	{\small \begin{align}
			{\text{P6}}:\mathop {\max }\limits_{\bf{f}}\ &B\sum\limits_{k = 1}^K {{\upsilon _k}{{\log }_2}\big(1 + \frac{{{{\left| {{{\mathbf{f}}^{\mathrm{H}}}{{\mathbf{A}}_k}{{\mathbf{w}}_k}} \right|}^2}}}{{\sum\limits_{i = 1,i \ne k}^K {{{\left| {{{\mathbf{f}}^{\mathrm{H}}}{{\mathbf{A}}_k}{{\mathbf{w}}_i}} \right|}^2}} + \sigma _k^2}}\big)}, \\
			\mathrm{s.t.}\ &\frac{{{{\left| {{{\mathbf{f}}^{\mathrm{H}}}{{\mathbf{A}}_k}{{\mathbf{w}}_k}} \right|}^2}}}{{\sum\limits_{i = 1,i \ne k}^K {{{\left| {{{\mathbf{f}}^{\mathrm{H}}}{{\mathbf{A}}_k}{{\mathbf{w}}_i}} \right|}^2}} + \sigma _k^2}} \ge r_k^{\mathrm{EIA}},\forall k, \\
			&\Omega^{\mathrm{EIA}} \le \frac{{{{\left| {{{\mathbf{f}}^{\mathrm{H}}}{{\mathbf{A}}_k}{{\mathbf{w}}_c}} \right|}^2}}}{{\sum\limits_{i = 1}^K {{{\left| {{{\mathbf{f}}^{\mathrm{H}}}{{\mathbf{A}}_k}{{\mathbf{w}}_i}} \right|}^2}} + \sigma _k^2}},\forall k, \\
			&\left| {f _n^l} \right| = 1,\forall n,l.
	\end{align}}%
\end{subequations}
where $ r_k^{\mathrm{EIA}} = {2^{{R_{k}} - C_k^{\mathrm{EIA}}}} - 1 $ and $ \Omega^{\mathrm{EIA}} = {2^{\sum\limits_{k = 1}^K {C_k^{\mathrm{EIA}}} }} - 1 $.

Denote $ {\mathbf{u}} = [{\textbf{f}_1};...;{\textbf{f}_n};...;{\textbf{f}_N}] $, $ {{\mathbf{a}}_{k,i}} = {{\mathbf{A}}_k}(1:LN,:){{\mathbf{w}}_i} $ and $ {\alpha _{k,i}} = {\mathbf{g}}_k^H{{\mathbf{w}}_i} $, where $ {{\mathbf{A}}_k}(1:LN,:) $ represents the submatrices containing the elements from the $ 1 $-th row to the $ LN $-th row of $ {{\mathbf{A}}_k} $. Hence, problem P6 is rewritten as 
\begin{subequations}
	{\footnotesize \begin{align}
			{\text{P7}}: \mathop {\max }\limits_{\mathbf{u}}&\ B\sum\limits_{k = 1}^K {{\upsilon _k}{{\log }_2}\big(1 + \frac{{{{\left| {{{\mathbf{u}}^{\mathrm{H}}}{{\mathbf{a}}_{k,k}} + {\alpha _{k,k}}} \right|}^2}}}{{\sum\limits_{i = 1,i \ne k}^K {{{\left| {{{\mathbf{u}}^{\mathrm{H}}}{{\mathbf{a}}_{k,i}} + {\alpha _{k,i}}} \right|}^2}} + \sigma _k^2}}\big)}, \\
			\mathrm{s.t.}\ &\frac{{{{\left| {{{\mathbf{u}}^{\mathrm{H}}}{{\mathbf{a}}_{k,k}} + {\alpha _{k,k}}} \right|}^2}}}{{\sum\limits_{i = 1,i \ne k}^K {{{\left| {{{\mathbf{u}}^{\mathrm{H}}}{{\mathbf{a}}_{k,i}} + {\alpha _{k,i}}} \right|}^2}}  + \sigma _k^2}} \ge r_k^{\mathrm{EIA}},\forall k, \\
			&\Omega^{\mathrm{EIA}} \le \frac{{{{\left| {{{\mathbf{u}}^{\mathrm{H}}}{{\mathbf{a}}_{k,c}} + {\alpha _{k,c}}} \right|}^2}}}{{\sum\limits_{i = 1}^K {{{\left| {{{\mathbf{u}}^{\mathrm{H}}}{{\mathbf{a}}_{k,i}} + {\alpha _{k,i}}} \right|}^2}}  + \sigma _k^2}},\forall k, \\
			&\left| {{{\mathbf{u}}_i}} \right| = 1,i = 1,...,LN.
	\end{align}}%
\end{subequations}

Let $ \overline {\mathbf{u}} = [{\mathbf{u}};1] $ and {\small $ {\bm{\Gamma}} _{k,i}^\varepsilon = [{{\mathbf{a}}_{k,i}}{\mathbf{a}}_{k,i}^{\mathrm{H}},{{\mathbf{a}}_{k,i}}\alpha _{k,i}^{\mathrm{H}};{\alpha _{k,i}}{\mathbf{a}}_{k,i}^{\mathrm{H}},0] $}, we can obtain 
\begin{align}
	{\overline {\mathbf{u}} ^{\mathrm{H}}}{\bm{\Gamma}} _{k,i}^\varepsilon \overline {\mathbf{u}}  = {\text{Tr}}({\bm{\Gamma}} _{k,i}^\varepsilon \overline {\mathbf{u}}\, {\overline {\mathbf{u}} ^{\mathrm{H}}}),
\end{align} 

\noindent Denote ${\mathbf{V}} = \overline{\mathbf{u}}\, {\overline{\mathbf{u}} ^{\mathrm{H}}}$, where ${\mathbf{V}} \succcurlyeq {\mathbf{0}}$ and ${\text{Rank}}({\mathbf{V}}) = 1$. Through the utilization of the SDR method to eliminate the rank-one constraint, problem P7 will be revised as
\begin{subequations}
	{\small \begin{align}
			{\text{P8}}: \mathop {\max }\limits_{\mathbf{V}}\ &B\sum\limits_{k = 1}^K {{\upsilon _k}{{\log }_2}\big(1 + \frac{{{\text{Tr}}({\bm{\Gamma}} _{k,k}^\varepsilon {\mathbf{V}}) + {{\left| {{\alpha _{k,k}}} \right|}^2}}}{{\sum\limits_{i = 1,i \ne k}^K {{\text{Tr}}({\bm{\Gamma}} _{k,i}^\varepsilon {\mathbf{V}}) + {{\left| {{\alpha _{k,i}}} \right|}^2}} + \sigma _k^2}}\big)}, \\
			\mathrm{s.t.}\ &\frac{{{\text{Tr}}({\bm{\Gamma}} _{k,k}^\varepsilon {\mathbf{V}}) + {{\left| {{\alpha _{k,k}}} \right|}^2}}}{{\sum\limits_{i = 1,i \ne k}^K {{\text{Tr}}({\bm{\Gamma}} _{k,i}^\varepsilon {\mathbf{V}}) + {{\left| {{\alpha _{k,i}}} \right|}^2}} + \sigma _k^2}} \ge r_k^{\mathrm{EIA}},\forall k,\label{eq:41b} \\
			&\Omega^{\mathrm{EIA}} \le \frac{{{\text{Tr}}({\bm{\Gamma}} _{k,c}^\varepsilon {\mathbf{V}}) + {{\left| {{\alpha _{k,c}}} \right|}^2}}}{{\sum\limits_{i = 1}^K {{\text{Tr}}({\bm{\Gamma}} _{k,i}^\varepsilon {\mathbf{V}}) + {{\left| {{\alpha _{k,i}}} \right|}^2}} + \sigma _k^2}},\forall k, \\
			&{\mathbf{V}} \succcurlyeq {\mathbf{0}},\\
			&{{\mathbf{V}}_{i,i}} = 1,i = 1,...,(LN + 1).\label{eq:41e}
	\end{align}}
\end{subequations}

According to the properties of the logarithmic functions, problem P8 can be rewritten as 
\begin{subequations}
	\begin{align}
		{\text{P9}}: \mathop {\max }\limits_{\mathbf{V}}\ &B\sum\limits_{k = 1}^K {{\upsilon _k}} ({P_k}({\mathbf{V}}) + {Z_k}({\mathbf{V}})),\\
		\mathrm{s.t.} \ &\text{(\ref{eq:41b})} - \text{(\ref{eq:41e})}.
	\end{align}
\end{subequations}
where 
{\small \begin{align}
		&{P_k}({\mathbf{V}}) = {\log _2}\big(\sum\limits_{i = 1}^K {{\text{Tr}}({\bm{\Gamma}} _{k,i}^\varepsilon {\mathbf{V}}) + {{\left| {{\alpha _{k,i}}} \right|}^2}} + \sigma _k^2\big),\forall k, \\
		&{Z_k}({\mathbf{V}}) = - {\log _2}\big(\sum\limits_{i = 1,i \ne k}^K {{\text{Tr}}({\bm{\Gamma}} _{k,i}^\varepsilon {\mathbf{V}}) + {{\left| {{\alpha _{k,i}}} \right|}^2}} + \sigma _k^2\big),\forall k.
\end{align}}%

\noindent By applying Theorem \ref{Theorem:1}, the lower bound of {\small $ {Z_k}({\mathbf{V}}) $} is 
{\footnotesize \begin{align}
	Z_k^{L,(t)}({{\mathbf{V}}^{(t)}}) &= \mathop {\max }\limits_{\varrho _k^{(t)} > 0} - \frac{{\varrho _k^{(t)}\big(\sum\limits_{i = 1,i \ne k}^K {{\text{Tr}}({\bm{\Gamma}} _{k,i}^\varepsilon {{\mathbf{V}}^{(t)}}) + {{\left| {{\alpha _{k,i}}} \right|}^2}}  + \sigma _k^2\big)}}{{\ln 2}} \notag \\
	& + {\log _2}\varrho _k^{(t)} + (1/\ln 2),\label{eq:45}
\end{align}}%
where 
{\small \begin{align}
	\varrho _k^{(t)} = {\big(\sum\limits_{i = 1,i \ne k}^K {{\text{Tr}}({\bm{\Gamma}} _{k,i}^\varepsilon {{\mathbf{V}}^{(t - 1)}}) + {{\left| {{\alpha _{k,i}}} \right|}^2}}  + \sigma _k^2\big)^{ - 1}}.\label{eq:46}
\end{align}}%

\noindent Solving problem P10 in the $t$-th iteration yields a lower bound solution for optimization problem P9
\begin{subequations}
	\begin{align}
			&{\text{P10}}: \mathop {\max }\limits_{{{\mathbf{V}}^{(t)}}}\ B\sum\limits_{k = 1}^K {{\upsilon _k}\big({P_k}({{\mathbf{V}}^{(t)}}) + } Z_k^{L,(t)}({{\mathbf{V}}^{(t)}})\big), \\
			&\mathrm{s.t.}\ \frac{{{\text{Tr}}({\bm{\Gamma}} _{k,k}^\varepsilon {{\mathbf{V}}^{(t)}}) + {{\left| {{\alpha _{k,k}}} \right|}^2}}}{{\sum\limits_{i = 1,i \ne k}^K {{\text{Tr}}({\bm{\Gamma}} _{k,i}^\varepsilon {{\mathbf{V}}^{(t)}}) + {{\left| {{\alpha _{k,i}}} \right|}^2}} + \sigma _k^2}} \ge r_k^{\mathrm{EIA}},\forall k, \\
			&\qquad \Omega^{\mathrm{EIA}} \le \frac{{{\text{Tr}}({\bm{\Gamma}} _{k,c}^\varepsilon {{\mathbf{V}}^{(t)}}) + {{\left| {{\alpha _{k,c}}} \right|}^2}}}{{\sum\limits_{i = 1}^K {{\text{Tr}}({\bm{\Gamma}} _{k,i}^\varepsilon {{\mathbf{V}}^{(t)}}) + {{\left| {{\alpha _{k,i}}} \right|}^2}} + \sigma _k^2}},\forall k, \\
			&\qquad {{\mathbf{V}}^{(t)}} \succcurlyeq {\mathbf{0}},\\
			&\qquad {\mathbf{V}}_{i,i}^{(t)} = 1,i = 1,...,(LN + 1).
	\end{align}
\end{subequations}

The problem P10 has been shown to be a convex problem, and thus can be solved using the CVX toolbox.

\begin{algorithm}[t] 
	\caption{AO Algorithm based on Dinkelbach for Solving Problem P3.} 
	\begin{algorithmic}[1]
		\STATE
		\textbf{Initialize:} {\small $ {\bf{W}}^{(0)} $}, $ {{\bf{f}}^{(0)}} $, and $ \rho_1^{(0)} = 0 $, iteration index $ t=1 $ and accuracy threshold $ \varepsilon > 0 $.
		\STATE
		\textbf{Repeat:}
		\STATE
		In the $ t $-th iteration, with the given $ \rho_1^{(t-1)} $, {\small $ {\bf{W}}^{(t-1)} $} and $ {{\bf{f}}^{(t-1)}} $, solve the problem P5 to obtain {\small $ {\mathbf{W}}^{(t)} $} and {\small $ ({\mathbf{C}}^{\mathrm{EIA}})^{(t)} $}, and decompose $ {\mathbf{w}}^{(t)} $ from {\small $ {\mathbf{W}}^{(t)} $}.
		\STATE
		For given $ {\mathbf{w}}^{(t)} $, {\small $ ({\mathbf{C}}^{\mathrm{EIA}})^{(t)} $}, $ \rho_1^{(t-1)} $, and $ {{\bf{f}}^{(t-1)}} $, solve the problem P10 to obtain {\small $ {\mathbf{V}}^{(t)} $}, and decompose $ {{\bf{f}}^{(t)}} $ from {\small $ {\mathbf{V}}^{(t)} $}.
		\STATE
		Update {\small $ \rho_1^{(t)} = \frac{B\sum\limits_{k = 1}^K {{\upsilon _k}}R_k^{\mathrm{EIA}}({\mathbf{w}}^{(t)},({\mathbf{C}}^{\mathrm{EIA}})^{(t)}, {{\bf{f}}^{(t)}})}{{\cal P}_{\mathrm{total}}^{{\mathrm{EIA}}}({\mathbf{w}}^{(t)})} $}. 
		\STATE
		Update $ t = t + 1 $.
		\STATE
		\textbf{Until:} $ \left|\rho_1^{(t)}-\rho_1^{(t-1)}\right| \le \varepsilon $.
		\label{algorithm1} 
	\end{algorithmic} 
\end{algorithm}

\subsection{Convergence Analysis}
The overall proposed algorithm is outlined in Algorithm 1. The convergence of Algorithm 1 can be established through the theorem 3.

\begin{Theorem}\label{Theorem:3}
	The objective function value of problem P1 monotonically increases over the iterations with the implementation of Algorithm 1.
\end{Theorem}

\begin{Proof}
	Please refer to the Appendix B.
        $\hfill \square$
\end{Proof}

\subsection{Computational Complexity}

As stated in \cite{9133120}, an optimization algorithm's complexity is dependent on the quantity of variables and constraints present in the problem. Therefore, the computational complexity of Algorithm 1 is predominantly dictated by the resolution of problems P5 and P10. Let the iteration count of the AO-based algorithm (i.e., Algorithm 1) be denoted as $I_{\mathrm{AO}_1}$. We proceed to examine the computational complexity associated with problem P5 and P10 individually.

For problem P5, the iteration number, as outlined in \cite{6891348,9133120}, is denoted by $I_{\mathrm{iter}_1} = \sqrt{(K + 1)M + \ell_1}$, where ${\ell_1} = 3K + 1$. Within each iteration, the computational complexity is approximately as $I_1 = \mathcal{O}\left({n_1}((K + 1){M^3} + {\ell_1}) + n_1^2((K + 1){M^2} + {\ell_1}) + n_1^3\right)$, where ${n_1} = (K + 1){M^2} + K$. Hence, the computational complexity for solving problem P5 is $ {\cal O}(I_{\mathrm{iter}_1}I_{1}) $

The complexity associated with solving problem P10 is given by $ \mathcal{O}(I_{\mathrm{iter}_2}I_{2}) $, where $ I_{\mathrm{iter}_2} = \sqrt {(LN + 1) + {\ell _2}} $ and $ I_2 = \mathcal{O}({n_2}({(LN + 1)^3} + {\ell _2}) + n_2^2({(LN + 1)^2} + {\ell _2}) + n_2^3) $ with $ {\ell _2} = 2K + LN + 1 $ and $ {n_2} = {(LN + 1)^2} $.

Therefore, the computational complexity of Algorithm 1 is $ {\cal O}( I_{\mathrm{AO}_1}(I_{\mathrm{iter}_1}I_{1} + I_{\mathrm{iter}_2}I_2)) $.

\section {Problem Formulation and Solution for the OIA Scheme}

This section studies the weighted EE maximization problem under the OIA scheme, involving the joint optimization of the AP transmit beamforming, the IRS-user association coefficients, the IRS phase shift matrices, and the user common rates. Denoted as {\small $ {{\mathbf{C}}^{\mathrm{OIA}}} = [C_1^{\mathrm{OIA}},...,C_K^{\mathrm{OIA}}] $}, the optimization problem is articulated as follows
\begin{subequations}
	{\small \begin{align}
			{\text{P11}}:\mathop {\max }\limits_{{\mathbf{w}},{\bm{\theta}}_n, {{\mathbf{v}}_{n,k}},{{\mathbf{C}}^{\mathrm{OIA}}}} &\frac{{B\sum\limits_{n = 1}^N {\sum\limits_{k = 1}^K {{\upsilon _k}{\varpi _{n,k}}R_{n,k}^{\mathrm{OIA}}} } }}{{\cal P}_{\mathrm{total}}^{\mathrm{OIA}}},\label{eq:48a} \\
			\mathrm{s.t.}\, \qquad&{\left\| {{{\mathbf{w}}_c}} \right\|^2} + \sum\limits_{k = 1}^K {{{\left\| {{{\mathbf{w}}_k}} \right\|}^2}} \le {P_{\max }},\label{eq:48b} \\
			&\sum\limits_{n = 1}^N {{\varpi _{n,k}}}R_{n,k}^{\mathrm{OIA}} \ge {R_{k}},\forall k,\label{eq:48c} \\
			&\sum\limits_{k = 1}^K {C_k^{\mathrm{OIA}}} \le \sum\limits_{n = 1}^N {{\varpi _{n,k}}}R_{c,n,k}^{\mathrm{OIA}}, \forall k, \label{eq:48d} \\
			&C_k^{\mathrm{OIA}} \ge 0,\forall k,\label{eq:48e}\\
			&{\varpi _{n,k}} \in \{0,1\},\forall n,k,\label{eq:48f} \\
			&\sum\limits_{k = 1}^K {\varpi _{n,k}} \le a,\forall n,\label{eq:48g} \\
			&\sum\limits_{n = 1}^N {{\varpi _{n,k}}} = 1,\forall k,\label{eq:48h}\\
			&\left| {\theta _n^l} \right| = 1,\forall n,l,\label{eq:48i}
	\end{align}}%
\end{subequations}
where (\ref{eq:48b}) represents the power constraint at the AP, constraint (\ref{eq:48c}) denotes the minimal rate threshold of user $ k $, and
constraint (\ref{eq:48d}) guarantees that the sum common rate of all users remains within the bounds of the system common rate. Constraint (\ref{eq:48f}) imposes a binary constraint on the association coefficient between the IRSs and users. Additionally, constraint (\ref{eq:48g}) limits the number of users that associated with the same IRS cannot exceed $ a $, and constraint (\ref{eq:48h}) restricts each user to be associated with only one IRS. (\ref{eq:48i}) corresponds to the phase shift constraint at the IRSs.

It is evident that problem P11 is a MINLP problem attributed to the non-convex fractional objective function, the non-convex constraints (\ref{eq:48c}), (\ref{eq:48d}), and (\ref{eq:48i}) as well as the binary constraints (\ref{eq:48f})-(\ref{eq:48h}).

\subsection{Problem Transformation}

By applying the Dinkelbach method on the objective function (\ref{eq:48a}), the problem P11 can be transformed into 
\begin{subequations}
	\begin{align}
			&{\text{P}}12:\mathop {\max }\limits_{{\mathbf{w}},{\bm{\theta}}_n, {{\mathbf{v}}_{n,k}},{{\mathbf{C}}^{\mathrm{OIA}}}}\ F({\rho _2}), \\
			&\mathrm{s.t.}\quad\text{(\ref{eq:48b})} -\text{(\ref{eq:48i})},
	\end{align}
\end{subequations}
where $ {\rho _2} $ is an auxiliary variable and
{\footnotesize \begin{align}
		F({\rho _2}) &= B\sum\limits_{n = 1}^N \sum\limits_{k = 1}^K {\upsilon _k}{\varpi _{n,k}}\big(C_k^{\mathrm{OIA}} \notag \\
		& + {{\log }_2}(1 + \frac{{{{\left| {\widetilde {\mathbf{H}}_{n,k}^{\mathrm{H}}{{\mathbf{w}}_k}} \right|}^2}}}{{\sum\limits_{i = 1,i \ne k}^K {{{\left| {\widetilde {\mathbf{H}}_{n,k}^{\mathrm{H}}{{\mathbf{w}}_i}} \right|}^2}} + \delta _k^2}})\big) - {\rho _2}{\cal P}_{\mathrm{total}}^{\mathrm{OIA}}.\label{eq:50}
\end{align}}%

\noindent Denote $ {{\mathbf{E}}_{n,k}} = [{\text{diag}}({\mathbf{h}}_{n,k}^{\mathrm{H}}){{\mathbf{G}}_n};{\mathbf{g}}_k^{\mathrm{H}}] $ and $ \widetilde {\mathbf{f}}_n = [{{\mathbf{f}}_n};1] $, then
{\footnotesize \begin{align}
	\widetilde {\mathbf{H}}_{n,k}^{\mathrm{H}}{{\mathbf{w}}_\tau } = \widetilde{\mathbf{f}}_n^{\mathrm{H}}{{\mathbf{E}}_{n,k}}{{\mathbf{w}}_\tau },\tau  = c,1,2,...,K,\label{eq:51}
\end{align}}

\noindent By substituting (\ref{eq:51}) into (\ref{eq:50}), $ F({\rho _2}) $ can be rewritten as
{\footnotesize \begin{align}
		{F^{'}}({\rho _2}) &= B\sum\limits_{n = 1}^N \sum\limits_{k = 1}^K {\upsilon _k}{\varpi _{n,k}}\big(C_k^{\mathrm{OIA}}\notag \\
		& + {{\log }_2}(1 + \frac{{{{\left| {\widetilde {\mathbf{f}}_n^{\mathrm{H}}{{\mathbf{E}}_{n,k}}{{\mathbf{w}}_k}} \right|}^2}}}{{\sum\limits_{i = 1,i \ne k}^K {{{\left| {\widetilde {\mathbf{f}}_n^{\mathrm{H}}{{\mathbf{E}}_{n,k}}{{\mathbf{w}}_i}} \right|}^2}}  + \delta _k^2}})\big) - {\rho _2}{\cal P}_{\mathrm{total}}^{\mathrm{OIA}},\label{eq:52}
\end{align}}

\noindent Then, substituting (\ref{eq:51}) and (\ref{eq:52}) into problem P12, P12 can be transformed into 
\begin{subequations}
	{\footnotesize \begin{align}
			&{\text{P}}13:\mathop {\max }\limits_{{\mathbf{w}},{\bf{\widetilde f}}_n, {{\mathbf{v}}_{n,k}}, {{\mathbf{C}}^{\mathrm{OIA}}}}\ {F^{'}}({\rho _2}), \\
			&\mathrm{s.t.}\quad \text{(\ref{eq:48b})}, \text{(\ref{eq:48e})}-\text{(\ref{eq:48h})},\\
			&\sum\limits_{n = 1}^N {{\varpi _{n,k}}}\big(C_k^{\mathrm{OIA}} + {\log _2}(1 + \frac{{{{\left| { \widetilde {\mathbf{f}}_n^{\mathrm{H}}{{\mathbf{E}}_{n,k}}{{\mathbf{w}}_k}} \right|}^2}}}{{\sum\limits_{i = 1,i \ne k}^K {{{\left| { \widetilde {\mathbf{f}}_n^{\mathrm{H}}{{\mathbf{E}}_{n,k}}{{\mathbf{w}}_i}} \right|}^2}} + \delta _k^2}})\big) \ge {R_k},\forall n,k,\label{eq:53c} \\
			&\sum\limits_{k = 1}^K {C_k^{\mathrm{OIA}} \le } \sum\limits_{n = 1}^N {{\varpi _{n,k}}}{\log _2}\big(1 + \frac{{{{\left| { \widetilde {\mathbf{f}}_n^{\mathrm{H}}{{\mathbf{E}}_{n,k}}{{\mathbf{w}}_c}} \right|}^2}}}{{\sum\limits_{i = 1}^K {{{\left| { \widetilde {\mathbf{f}}_n^{\mathrm{H}}{{\mathbf{E}}_{n,k}}{{\mathbf{w}}_i}} \right|}^2}}  + \delta _k^2}}\big),\forall n,k,\label{eq:53d} \\
			&\left| {f _n^l} \right| = 1,\forall n,l.\label{eq:53e}
	\end{align}}%
\end{subequations}

Obviously, problem P13 is non-convex and is tackled by decomposing it into three subproblems. These subproblems optimize variables $ \{{\mathbf{v}}_{n,k}\} $, {\small $ \{{\mathbf{w}}, {\mathbf{C}}^{\mathrm{OIA}}\} $}, and $ {\widetilde {\bf f}}_n $ alternately with fixed other variables.

\subsection{IRS Association Optimization}

For the given {\small $ \{{\mathbf{w}}, {\mathbf{C}}^{\mathrm{OIA}}\} $} and $ {\widetilde {\bf f}}_n $, problem P13 can be simplified as follows by removing the irrelevant terms\footnote{It is worth noting that although constraints (\ref{eq:53c}) and (\ref{eq:53d}) contain the variable $ {\varpi _{n,k}}$, they do not affect the optimal solution. Because constraints (\ref{eq:53c}) and (\ref{eq:53d}) are consistent with the optimization of the objective function, which aims to select the IRS that maximizes the total achievable rate for each user, therefore, they can be removed from problem P14.}
\begin{subequations}
	{\footnotesize \begin{align}
		{\text{P}}14:\mathop {\max }\limits_{{{\mathbf{v}}_{n,k}}}\ &B\sum\limits_{n = 1}^N {\sum\limits_{k = 1}^K {{\upsilon _k}{\varpi _{n,k}}R_{n,k}^{\mathrm{OIA}}} } - {\rho _2}{\cal P}_{\mathrm{total}}^{\mathrm{OIA}}, \\
		\mathrm{s.t.}\ \ &{\varpi _{n,k}} \in \{0,1\},\forall n,k,\label{eq:54b} \\
		&\sum\limits_{k = 1}^K {{\varpi _{n,k}}}  \le a,\forall n,\label{eq:54c} \\
		&\sum\limits_{n = 1}^N {{\varpi _{n,k}}} = 1,\forall k.\label{eq:54d}
	\end{align}}%
\end{subequations}
Constrained by the binary boundary (\ref{eq:54b}), directly solving problem P14 proves to be ineffective. We firstly denote the optimal objective function value of problem P14 as $ \chi ^\star $.

The BnB method is proved with great potential as a comprehensive and methodic method to tackle complex optimization problems in a discrete and combinatorial setting \cite{boyd2007branch}. And the application of BnB method in effectively addressing the issue of extensively resource allocation has been extensively investigated \cite{7934461,5200968,Xu2022}.

\subsubsection{Lower and Upper Bounds}
By introducing continuous auxilary variables $ \{\omega _{n,k}\} $ to represent the relaxed version of the binary variables $ {\varpi _{n,k}} $ and removing the irrelevant terms in P14, the relaxation problem becomes a linear problem,
\begin{subequations}
	{\small \begin{align}
		{\text{P}}15:\mathop {\max }\limits_{\left\{ {{\omega _{n,k}}} \right\}}\ &F_r({\omega _{n,k}}), \\
		\mathrm{s.t.}\ \ \ &0 \le {\omega _{n,k}} \le 1,\forall n,k,\\
		&\sum\limits_{k = 1}^K {{\omega _{n,k}}} \le a,\forall n,\label{eq:55c} \\
		&\sum\limits_{n = 1}^N {{\omega _{n,k}}} = 1,\forall k,\label{eq:55d}
	\end{align}}%
\end{subequations}
where 
{\footnotesize \begin{align}
	F_r({\omega _{n,k}}) &= B\sum\limits_{n = 1}^N {\sum\limits_{k = 1}^K {{\upsilon _k}{\omega _{n,k}}R_{n,k}^{\mathrm{OIA}}} }  - \rho_2\sum\limits_{n = 1}^N \max \{\omega _{n,k},\forall k\} L {P}(b). 
\end{align}}%
Notably, problem P15 can be solved directly. The optimal objective function value of P15, denoted as $ U^{(0)} $, is an upper bound of $ \chi ^\star $. We can also obtain the lower bound of $ \chi ^\star $ based on the solution of problem P15. Specifically, assuming that the optimal solution obtained by solving the relaxed optimization problem P15 is $ \{\omega _{n,k}^\star\} $, and $ \omega _{n,k}^\star \in [0,1],\forall n,k $. Then, by rounding $ \omega _{n,k}^\star $ to $ 0 $ or $ 1 $, we can construct a binary solution for problem P14. In particular, for $ \forall k $, the variable $ \omega _{n,k}^\star $ is rounded by
\begin{align}
	\omega _{n,k}^\star = \left\{ \begin{array}{l}
		1,\ n = \arg \mathop {\max }\limits_n \omega _{n,k}^\star\\
		0,\ \text{otherwise}
	\end{array} \right.\label{eq:56}
\end{align}
We use $ \{\widetilde \omega _{n,k}\} $ to represent the rounded solution. Then substitute $ \{\widetilde \omega _{n,k}\} $ into problem P14 to obtain the corresponding lower bound of the objective function value denoted as $ L^{(0)} $.

\subsubsection{Partitioning Rule and Branching Strategy}
Given the iteration index denoted by superscript $ (j) $ for the optimization variables, we commence by selecting the node linked to the largest Euclidean distance\footnote{The Euclidean distance represents the distance between the node and the current optimal solution. Choosing the node with the maximum distance helps to quickly identify nodes that are not better than the current optimal solution, thus allowing for earlier pruning and reducing the size of the search space. This can save computational resources and time \cite{Xu2022}.}
between $ \omega_{n,k} $ and its rounded counterpart $ \widetilde{\omega}_{n,k} $. More precisely, in the $ j $-th iteration, the root node to branch is determined by the index $ (n^*, k^*) $, where $ (n^*, k^*) $ is computed as
{\footnotesize \begin{align}
	({n^*},{k^*}) = \arg \mathop {\max }\limits_{n,k} \left| {(\omega _{n,k}^{\star})^{(j)} - \widetilde \omega _{n,k}^{(j)}} \right|,\label{eq:57}
\end{align}}%
then we can form two problems: The first problem is
\begin{subequations}
	{\footnotesize \begin{align}
		{{\cal{P}}0}:\mathop {\max }\limits_{\left\{ {{\varpi _{n,k}}} \right\}}\ &\sum\limits_{n = 1}^N \big(B{\sum\limits_{k = 1}^K {{\upsilon _k}\varpi _{n,k}^{(j)}R_{n,k}^{\mathrm{OIA}}} - \rho_2 \max \{\varpi _{n,k}^{(j)},\forall k\} L {P}(b)}\big), \\
		\mathrm{s.t.}\ &\varpi _{n,k}^{(j)} \in \{ 0,1\} ,\forall n \in {\cal N} \backslash \{n^*\},k \in {\cal K} \backslash \{k^*\},\label{eq:58b} \\
		&\varpi _{{n^*},{k^*}}^{(j)} = 0,\\
		&\sum\limits_{k = 1}^K {{\varpi _{n,k}^{(j)}}} \le a,\forall n,\label{eq:54c} \\
		&\sum\limits_{n = 1}^N {{\varpi _{n,k}^{(j)}}} = 1,\forall k.\label{eq:54d}
	\end{align}}%
\end{subequations}
where sets $ \cal N $ and $ \cal K $ collect the indices of the IRSs and users, respectively. And the second problem is
\begin{figure}[t]
	\centering
	\includegraphics[width=1\linewidth,height=0.7\linewidth]{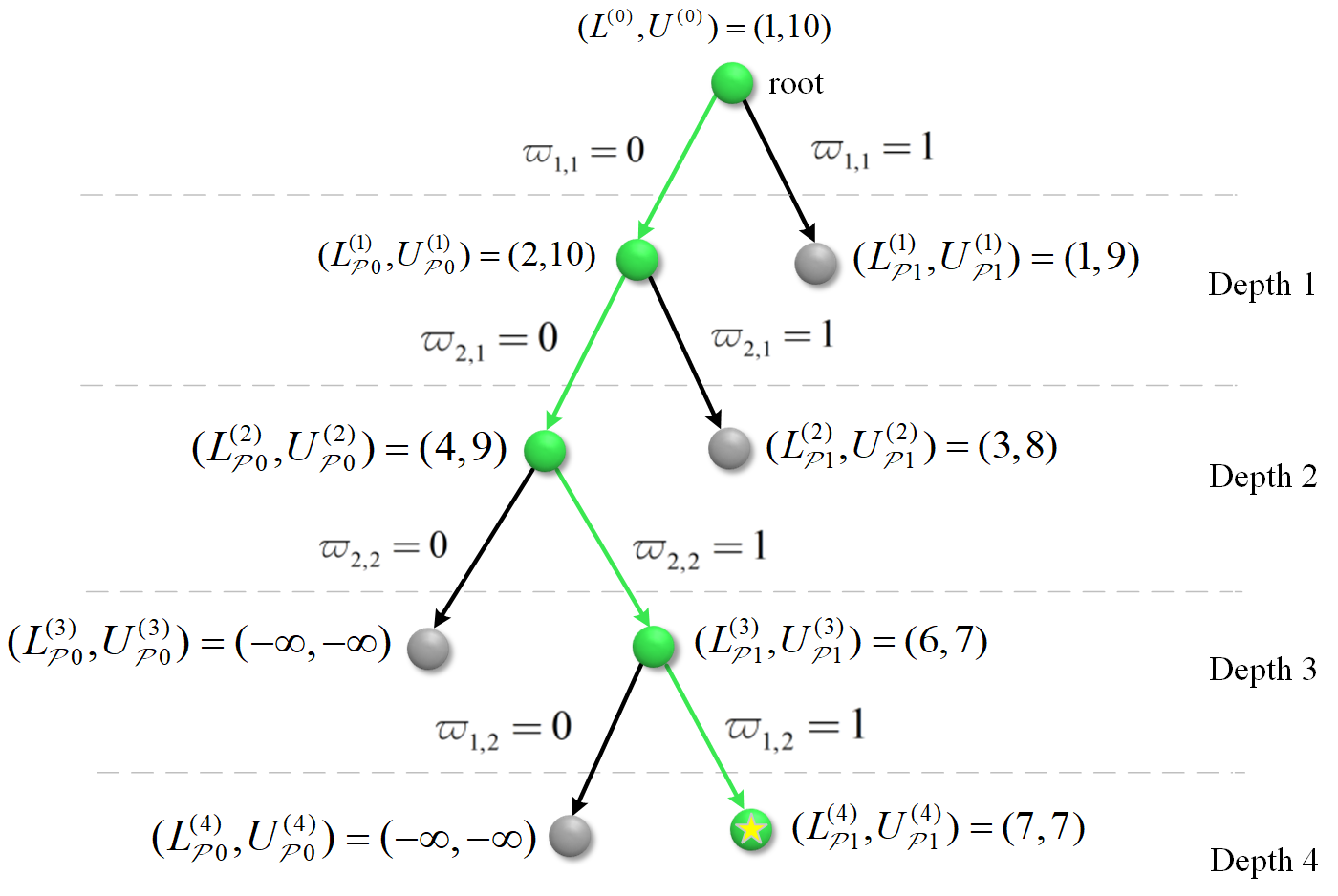}
	\caption{An example of the BnB search tree for $ N = 2 $ and $ K = 2 $.}
	\label{fig:bnb}
	\vspace{-1em}
\end{figure}
\begin{subequations}
	{\footnotesize \begin{align}
		{\cal{P}}1:\mathop {\max }\limits_{\left\{ {{\varpi _{n,k}}} \right\}}\ &\sum\limits_{n = 1}^N \big(B{\sum\limits_{k = 1}^K {{\upsilon _k}\varpi _{n,k}^{(j)}R_{n,k}^{\mathrm{OIA}}} - \rho_2 \max \{\varpi _{n,k}^{(j)},\forall k\} L {P}(b)}\big), \\
		\mathrm{s.t.}\ &\varpi _{n,k}^{(j)} \in \{ 0,1\} ,\forall n \in {\cal N} \backslash \{n^*\},k \in {\cal K} \backslash \{k^*\},\label{eq:59b} \\
		&\varpi _{{n^*},{k^*}}^{(j)} = 1,\\
		&\text{(\ref{eq:54c})},\text{(\ref{eq:54d})}.
	\end{align}}%
\end{subequations}

It can be observed that the problems $ {\cal{P}}0 $ and $ {\cal{P}}1 $ are non-convex attribute to the presence of constraints (\ref{eq:58b}) and (\ref{eq:59b}). In the $ j $-th iteration, by converting the binary variable $ {\varpi _{n,k}^{(j)}} $ into continuous variable between $ 0 $ and $ 1 $, we can successfully resolve the relaxed versions of both problem $ {\cal{P}}0 $ and $ {\cal{P}}1 $. Consequently, we acquire the optimal solutions along with their corresponding objective function values, denoted as $ U_{{\cal{P}}0}^{(j)} $ and $ U_{{\cal{P}}1}^{(j)} $, respectively. Subsequently, we utilize the obtained optimal solutions to problems $ {\cal{P}}0 $ and $ {\cal{P}}1 $ to derive their rounded solutions according to (\ref{eq:56}). By substituting the rounded solutions to problems $ {\cal{P}}0 $ and $ {\cal{P}}1 $ seperatively, the corresponding objective function values, denoted as $ L_{{\cal{P}}0}^{(j)} $ and $ L_{{\cal{P}}1}^{(j)} $ can be obtained. Therefore, $ (L_{{\cal{P}}0}^{(j)}, U_{{\cal{P}}0}^{(j)}) $ and $ (L_{{\cal{P}}1}^{(j)}, U_{{\cal{P}}1}^{(j)}) $ serve as the $ ( $\textit{lower bound, upper bound}$ ) $ respectively for problems $ {\cal{P}}0 $ and $ {\cal{P}}1 $ in the $ j $-th iteration.

Fig. \ref{fig:bnb} shows a BnB search tree example\footnote{It should be pointed out that the values in Fig. \ref{fig:bnb} are provided solely as examples. For specific numerical results, please refer to the Section V.} for $ N = 2 $ and $ K = 2 $, where the root node initially splits into two nodes corresponding to $ \varpi_{n,k} = 0 $ and $ \varpi_{n,k} = 1 $. Subsequently, the aforementioned process is iteratively repeated to obtain the upper and lower bounds for problem $ {\cal{P}}0 $ and $ {\cal{P}}1 $. Comparing the lower bounds of problems $ {\cal{P}}0 $ and $ {\cal{P}}1 $, select the node with higher lower bound\footnote{Notably, if the relaxed problem is infeasible, we let $ U_{{{\cal{P}}0}/{{\cal{P}}1}}^{(j)} = -\infty $ (and therefore $ L_{{{\cal{P}}0}/{{\cal{P}}1}}^{(j)} = -\infty $). In this case, this operation still has the capacity to maintain the algorithm's effective functioning.} as the new root node (represented by the green dot), while eliminating the unnecessary node (indicated by the grey dot). For the new root node, it is partitioned into two new nodes by selecting the new index according to (\ref{eq:57}), and then the lower and upper bounds for these two new nodes are calculated. The iteration carries on until the optimal solution is identified (indicated by the yellow star). The BnB approach guarantees termination within a finite number of iterations, owing to the bounded depth of the search tree and the finite nodes at each depth, determined by the number of variables.

Algorithm \ref{algorithm2} outlines the BnB-based approach for solving optimization problem P14.
 
\subsection{Common Rate Allocation and Transmit Beamforming Optimization}
For given $ \{{\mathbf{v}}_{n,k}\} $ and $ {\widetilde {\bf f}}_n $, let {\small $ {{\mathbf{F}}_{n,k}} = {\mathbf{E}}_{n,k}^{\mathrm{H}}{\widetilde {\mathbf{f}}_n}\widetilde {\mathbf{f}}_n^{\mathrm{H}}{{\mathbf{E}}_{n,k}} $},  the optimization problem P13 can be reduced as\footnote{In the preceding section, the optimal IRS associations have been obtained. Accordingly, we exclude the variable $ {\varpi _{n,k}} $ from the constraints (\ref{eq:61c}) and (\ref{eq:61d}). In which, the indices $ n $ and $ k $ will represent the indices where $ {\varpi _{n,k}} = 1 $.}
\begin{subequations}
	{\footnotesize \begin{align}
			&{\text{P}}16:\mathop {\max }\limits_{{\mathbf{W}},{{\mathbf{C}}^{\mathrm{OIA}}}} \ B\sum\limits_{n = 1}^N \sum\limits_{k = 1}^K {\upsilon _k}{\varpi _{n,k}}\big(C_k^{\mathrm{OIA}} + {I_{p,n,k}}({\mathbf{W}}) + {J_{p,n,k}}({\mathbf{W}})\big) \notag \\
			&\qquad\ - {\rho _2}\big({\text{Tr}}({{\mathbf{W}}_c}) + \sum\limits_{k = 1}^K {{\text{Tr}}({{\mathbf{W}}_k})} + {\Delta ^{\mathrm{OIA}}}\big), \label{eq:61a}\\
			&\ \mathrm{s.t.}\quad {\text{Tr}}({{\mathbf{W}}_c}) + \sum\limits_{k = 1}^K {{\text{Tr}}({{\mathbf{W}}_k})} \le {P_{\max }}, \\
			&\qquad\ \ C_k^{\mathrm{OIA}} + {I_{p,n,k}}({\mathbf{W}}) + {J_{p,n,k}}({\mathbf{W}}) \ge {R_k},\forall k, \label{eq:61c}\\
			&\qquad\ \ \sum\limits_{k = 1}^K {C_k^{\mathrm{OIA}} \le } {L_{c,n,k}}({\mathbf{W}}) + {M_{c,n,k}}({\mathbf{W}}),\forall k, \label{eq:61d}\\
			&\qquad\ \ C_k^{\mathrm{OIA}} \ge 0,\forall k, \\
			&\qquad\ \ {{\mathbf{W}}_\tau } \succcurlyeq {\mathbf{0}},\tau  = c, 1,...,K, \\
			&\qquad\ \ {\text{Rank(}}{{\mathbf{W}}_\tau }) = 1, \tau = c,1,...,K,
	\end{align}}%
\end{subequations}
where
{\scriptsize \begin{align}
		&{I_{p,n,k}}({\mathbf{W}}) = {\log _2}\big(\sum\limits_{i = 1}^K {{\text{Tr}}({{\mathbf{F}}_{n,k}}{{\mathbf{W}}_i})} + \delta _k^2\big),\\
		&{J_{p,n,k}}({\mathbf{W}}) = - {\log _2}\big(\sum\limits_{i = 1,i \ne k}^K {{\text{Tr}}({{\mathbf{F}}_{n,k}}{{\mathbf{W}}_i})} + \delta _k^2\big),\label{eq:62}\\
		&{L_{c,n,k}}({\mathbf{W}}) = {\log _2}\big(\sum\limits_{i = 1}^K {{\text{Tr}}({{\mathbf{F}}_{n,k}}{{\mathbf{W}}_i})} + {\text{Tr}}({{\mathbf{F}}_{n,k}}{{\mathbf{W}}_c}) + \delta _k^2\big),\\
		&{M_{c,n,k}}({\mathbf{W}}) = - {\log _2}\big(\sum\limits_{i = 1}^K {{\text{Tr}}({{\mathbf{F}}_{n,k}}{{\mathbf{W}}_i})}  + \delta _k^2\big),\label{eq:64}\\
		&{\Delta ^{\mathrm{OIA}}} = \sum\limits_{k = 1}^K {{P_{\mathrm{user},k}}} + {P_{\mathrm{AP}}} + \sum\limits_{n = 1}^N {\max \{\varpi _{n,k},\forall k\}} L {P}(b).
\end{align}}%
\begin{algorithm}[t] 
	\caption{BnB-based Algorithm for Solving Problem P14.}
	\begin{algorithmic}[1]
		\STATE
		\textbf{Initialize:} iteration index $ j = 1 $ and accuracy threshold $ 0 < \varepsilon \ll 1 $.
		\STATE
		With the parameters $ \rho_2 $, ${\mathbf{w}}$, ${\mathbf{C}}^{\mathrm{OIA}}$, and ${\widetilde {\bf f}}_n$ at hand, the problem P15 is solved to acquire the solution $ \{\omega_{n,k}^\star\}^{(0)} $ and the corresponding objective function value $ U^{(0)} $. Subsequently, the rounded solution $ \{\widetilde \omega _{n,k}\}^{(0)} $ is computed in accordance with equation (\ref{eq:56}), and substitute it into problem P14 to obtain the corresponding objective function value $ L^{(0)} $. Start by initializing the search tree with the inclusion of the root node linked to $ \{\omega _{n,k}\}^{(0)} $. The obtained $ (L^{(0)},U^{(0)}) $ corresponds to the lower and upper bounds of the root node.
		\STATE
		\textbf{Repeat:}
		\STATE
		In the $ j $-th iteration, proceed to branch the root node indexed as $ ({n^*,k^*}) $ using equation (\ref{eq:57}), and form two subproblems as $ {\cal P}0 $ and $ {\cal P}1 $.
		\STATE
		Solve the relaxed versions of the respective subproblems to determine the optimal solutions. Denote the resulting objective function values as $ U_{{\cal{P}}0}^{(j)} $ and $ U_{{\cal{P}}1}^{(j)} $, correspondingly.
		\STATE
		Based on the obtained solutions to the relaxed problems, we can obtain their rounded solutions according to (\ref{eq:56}).
		\STATE
		Solve the problem P14 by substituting the rounded solutions individually, and denote the corresponding objective function values as $ L_{{\cal{P}}0}^{(j)} $ and $ L_{{\cal{P}}1}^{(j)} $, respectively.
		\STATE
		For the lower and upper bounds combinations $ (L_{{\cal{P}}0}^{(j)}, U_{{\cal{P}}0}^{(j)}) $ and $ (L_{{\cal{P}}1}^{(j)}, U_{{\cal{P}}1}^{(j)}) $. Selecting the node with higher lower bound as the new root node, and update $ (L^{(j)},U^{(j)}) $ as the corresponding lower and upper bounds combination. 
		\STATE
		Update $ j = j+1 $.
		\STATE
		\textbf{Until:} $ U^{(j-1)} - L^{(j-1)} \le \varepsilon $.
	\end{algorithmic}
	\label{algorithm2} 
\end{algorithm}

\noindent Following Theorem 1, within the $ q $-th iteration, considering the specific point $ {\mathbf{W}}^{(q-1)} $ obtained in the preceding $ (q-1) $-th iteration, {\small $ {J_{p,n,k}}({\mathbf{W}}) $} can be replaced by
{\small \begin{align}
	J_{p,n,k}^{L,(q)}({{\mathbf{W}}^{(q)}}) &= \mathop {\max }\limits_{\varrho _{p,n.k}^{(q)}} - \frac{{\varrho _{p,n.k}^{(q)}\big(\sum\limits_{i = 1,i \ne k}^K {{\text{Tr}}({{\mathbf{F}}_{n,k}}{\mathbf{W}}_i^{(q)})} + \delta _k^2\big)}}{{\ln 2}} \notag \\ &+ {\log _2}\varrho _{p,n.k}^{(q)} + (1/\ln 2),\label{eq:66}
\end{align}}%
where 
{\small \begin{align}
	\varrho _{p,n.k}^{(q)} = {\big(\sum\limits_{i = 1,i \ne k}^K {{\text{Tr}}({{\mathbf{F}}_{n,k}}{\mathbf{W}}_i^{(q - 1)})} + \delta _k^2\big)^{ - 1}}.\label{eq:67}
\end{align}}%
\noindent Similarly, for ${M_{c,n,k}}({\mathbf{W}})$, its lower bound can be determined using a similar approach, and given by
{\small \begin{align}
	M_{c,n,k}^{L,(q)}({{\mathbf{W}}^{(q)}}) &= \mathop {\max }\limits_{\varrho _{c,n.k}^{(q)}} - \frac{{\varrho _{c,n.k}^{(q)}\big(\sum\limits_{i = 1}^K {{\text{Tr}}({{\mathbf{F}}_{n,k}}{\mathbf{W}}_i^{(q)})} + \delta _k^2\big)}}{{\ln 2}}\notag \\
	& + {\log _2}\varrho _{c,n.k}^{(q)} + (1/\ln 2),\label{eq:68}
\end{align}}%
where
{\small \begin{align}
	\varrho _{c,n.k}^{(q)} = {\big(\sum\limits_{i = 1}^K {{\text{Tr}}({{\mathbf{F}}_{n,k}}{\mathbf{W}}_i^{(q - 1)})} + \delta _k^2\big)^{ - 1}}.\label{eq:69}
\end{align}}%

\noindent Subsequently, the resolution of problem P16 is achieved by addressing problem P17 in the $ q $-th iteration.
\begin{subequations}
	{\footnotesize \begin{align}
		&{\text{P}}17:\mathop {\max }\limits_{{{\mathbf{W}}^{(q)}},{{\mathbf{C}}^{\mathrm{OIA}}}} B\sum\limits_{n = 1}^N \sum\limits_{k = 1}^K {\upsilon _k}{\varpi _{n,k}}\big(C_k^{\mathrm{OIA}} + {I_{p,n,k}}({{\mathbf{W}}^{(q)}}) + \notag \\
		&\qquad\qquad J_{p,n,k}^{L,(q)}({{\mathbf{W}}^{(q)}})\big) - {\rho _2}\big({\text{Tr}}({\mathbf{W}}_c^{(q)}) + \sum\limits_{k = 1}^K {{\text{Tr}}({\mathbf{W}}_k^{(q)})} + {\Delta ^{\mathrm{OIA}}}\big), \\
		&\mathrm{s.t.}\ {\text{Tr}}({\mathbf{W}}_c^{(q)}) + \sum\limits_{k = 1}^K {{\text{Tr}}({\mathbf{W}}_k^{(q)})}  \le {P_{\max }}, \\
		&\qquad C_k^{\mathrm{OIA}} + {I_{p,n,k}}({{\mathbf{W}}^{(q)}}) + J_{p,n,k}^{L,(q)}({{\mathbf{W}}^{(q)}}) \ge {R_k},\forall k, \\
		&\qquad \sum\limits_{k = 1}^K {C_k^{\mathrm{OIA}} \le }\ {L_{c,n,k}}({{\mathbf{W}}^{(q)}}) + M_{c,n,k}^{L,(q)}({{\mathbf{W}}^{(q)}}),\forall k, \\
		&\qquad C_k^{\mathrm{OIA}} \ge 0,\forall k, \\
		&\qquad {\mathbf{W}}_\tau ^{(q)} \succcurlyeq {\mathbf{0}}, \tau = c,1,...,K, \\
		&\qquad {\text{Rank(}}{\mathbf{W}}_\tau ^{(q)}{\text{) = 1,}}\tau = c,1,...,K,\label{eq:70g}
	\end{align}}
\end{subequations}
where the constraint (\ref{eq:70g}) can be eliminated using the SDR method, allowing for the direct solution of problem P17.

\subsection{Phase Shift Optimization}
For the optimization of the phase shift matrix of the selected IRSs, once $\{\mathbf{v}_{n,k}\}$ and $\{\mathbf{w},\mathbf{C}^{\mathrm{OIA}}\}$ are available, problem P13 can be reformulated as
\begin{subequations}
	{\small \begin{align}
		{\text{P}}18:\mathop {\max }\limits_{\widetilde {\bf{f}}_n}\ &B\sum\limits_{n = 1}^N {\sum\limits_{k = 1}^K {{\upsilon _k}{\varpi _{n,k}} {{\log }_2}\big(1 + \frac{{{{\left| {\widetilde {\mathbf{f}}_n^{\mathrm{H}}{{\mathbf{E}}_{n,k}}{{\mathbf{w}}_k}} \right|}^2}}}{{\sum\limits_{i = 1,i \ne k}^K {{{\left| {\widetilde {\mathbf{f}}_n^{\mathrm{H}}{{\mathbf{E}}_{n,k}}{{\mathbf{w}}_i}} \right|}^2}} + \delta _k^2}}\big)} }, \\
		\mathrm{s.t.}\ &\frac{{{{\left| { \widetilde {\mathbf{f}}_n^{\mathrm{H}}{{\mathbf{E}}_{n,k}}{{\mathbf{w}}_k}} \right|}^2}}}{{\sum\limits_{i = 1,i \ne k}^K {{{\left| { \widetilde {\mathbf{f}}_n^{\mathrm{H}}{{\mathbf{E}}_{n,k}}{{\mathbf{w}}_i}} \right|}^2}}  + \delta _k^2}} \ge r_k^{\mathrm{OIA}},\forall n,k, \\
		&\Omega^{\mathrm{OIA}} \le \frac{{{{\left| { \widetilde {\mathbf{f}}_n^{\mathrm{H}}{{\mathbf{E}}_{n,k}}{{\mathbf{w}}_c}} \right|}^2}}}{{\sum\limits_{i = 1}^K {{{\left| { \widetilde {\mathbf{f}}_n^{\mathrm{H}}{{\mathbf{E}}_{n,k}}{{\mathbf{w}}_i}} \right|}^2}}  + \delta _k^2}},\forall n,k, \\
		&\left| {f _n^l} \right| = 1,\forall n,l.
	\end{align}}%
\end{subequations}
where {\small $ r_k^{\mathrm{OIA}} = {2^{{R_{k}} - C_k^{\mathrm{OIA}}}} - 1 $} and {\small $ \Omega^{\mathrm{OIA}} = {2^{\sum\limits_{k = 1}^K {C_k^{\mathrm{OIA}}} }} - 1 $}.
Denote {\small $ {{\bf{b}}_{n,k,i}} = {{\mathbf{E}}_{n,k}}(1:L,:){{\mathbf{w}}_i} $} and {\small $ {\beta _{k,i}} = {\mathbf{g}}_k^{\mathrm{H}}{{\mathbf{w}}_i} $}, where $ {{\mathbf{E}}_{n,k}}(1:L,:) $ is the submatrix containing all elements from the $ 1 $-row to the $ L $-th row of $ {{\mathbf{E}}_{n,k}} $. This enables the reformulation of problem P18 as
\begin{subequations}
	{\footnotesize \begin{align}
			&{\text{P}}19:\mathop {\max }\limits_{{ { {\mathbf{f}}_n} }}\ 
			B\sum\limits_{n = 1}^N \sum\limits_{k = 1}^K {\upsilon _k}{\varpi _{n,k}}{{\log }_2}\big(1 + \notag \\
			 & \qquad\qquad\quad \frac{{{{\left| {{\mathbf{f}}_n^{\mathrm{H}}{{\mathbf{b}}_{n,k,k}} + {\beta _{k,k}}} \right|}^2}}}{{\sum\limits_{i = 1,i \ne k}^K {{{\left| {{\mathbf{f}}_n^{\mathrm{H}}{{\mathbf{b}}_{n,k,i}} + {\beta _{k,i}}} \right|}^2}} + \delta _k^2}}\big), \\
			&\mathrm{s.t.}\quad \frac{{{{\left| {{\mathbf{f}}_n^{\mathrm{H}}{{\mathbf{b}}_{n,k,k}} + {\beta _{k,k}}} \right|}^2}}}{{\sum\limits_{i = 1,i \ne k}^K {{{\left| {{\mathbf{f}}_n^{\mathrm{H}}{{\mathbf{b}}_{n,k,i}} + {\beta _{k,i}}} \right|}^2}} + \delta _k^2}} \ge r_k^{\mathrm{OIA}},\forall n,k, \\
			&\qquad\ {\Omega ^{\mathrm{OIA}}} \le \frac{{{{\left| {{\mathbf{f}}_n^{\mathrm{H}}{{\mathbf{b}}_{n,k,c}} + {\beta _{k,c}}} \right|}^2}}}{{\sum\limits_{i = 1}^K {{{\left| {{\mathbf{f}}_n^{\mathrm{H}}{{\mathbf{b}}_{n,k,i}} + {\beta _{k,i}}} \right|}^2}} + \delta _k^2}},\forall n,k, \\
			&\qquad\ \left| {f_n^l} \right| = 1,\forall n,l.
	\end{align}}%
\end{subequations}
\noindent Denote $ {\mathbf{\Gamma }}_{n,k,i}^\varepsilon  = [{{\mathbf{b}}_{n,k,i}}{\mathbf{b}}_{n,k,i}^{\mathrm{H}},{{\mathbf{b}}_{n,k,i}}\beta _{k,i}^{\mathrm{H}};{\beta _{k,i}}{\mathbf{b}}_{n,k,i}^{\mathrm{H}},0] $, then
{\small \begin{align}
	\widetilde {\mathbf{f}} _n^{\mathrm{H}}{\mathbf{\Gamma }}_{n,k,i}^\varepsilon {\widetilde {\mathbf{f}} _n} = {\text{Tr}}({\mathbf{\Gamma }}_{n,k,i}^\varepsilon {\widetilde {\mathbf{f}} _n}\widetilde {\mathbf{f}} _n^{\mathrm{H}}),
\end{align}}

\noindent Let ${\mathbf{F}}_n = {\widetilde{\mathbf{f}}_n}\widetilde{\mathbf{f}}_n^{\mathrm{H}}$, where ${\mathbf{F}}_n \succcurlyeq {\mathbf{0}}$ and ${\text{Rank}}({\mathbf{F}}_n) = 1$. Applying the SDR method to eliminate the rank-one constraint renders problem P19 equivalent to
\begin{subequations}
	{\footnotesize \begin{align}
			&{\text{P}}20:\mathop {\max }\limits_{{{{\mathbf{F}}_n}}}\ B\sum\limits_{n = 1}^N \sum\limits_{k = 1}^K {\upsilon _k}{\varpi _{n,k}}{{\log }_2}\big(1 + \notag \\ &\qquad\qquad\quad \frac{{{\text{Tr}}({\mathbf{\Gamma }}_{n,k,k}^\varepsilon {{\mathbf{F}}_n}) + {{\left| {{\beta _{k,k}}} \right|}^2}}}{{\sum\limits_{i = 1,i \ne k}^K {{\text{Tr}}({\mathbf{\Gamma }}_{n,k,i}^\varepsilon {{\mathbf{F}}_n}) + {{\left| {{\beta _{k,i}}} \right|}^2}}  + \delta _k^2}}\big),\label{eq:74a} \\
			&\mathrm{s.t.}\ \frac{{{\text{Tr}}({\mathbf{\Gamma }}_{n,k,k}^\varepsilon {{\mathbf{F}}_n}) + {{\left| {{\beta _{k,k}}} \right|}^2}}}{{\sum\limits_{i = 1,i \ne k}^K {{\text{Tr}}({\mathbf{\Gamma }}_{n,k,i}^\varepsilon {{\mathbf{F}}_n}) + {{\left| {{\beta _{k,i}}} \right|}^2}}  + \delta _k^2}} \ge r_k^{\mathrm{OIA}},\forall n,k, \\
			&\qquad{\Omega ^{\mathrm{OIA}}} \le \frac{{{\text{Tr}}({\mathbf{\Gamma }}_{n,k,c}^\varepsilon {{\mathbf{F}}_n}) + {{\left| {{\beta _{k,c}}} \right|}^2}}}{{\sum\limits_{i = 1}^K {{\text{Tr}}({\mathbf{\Gamma }}_{n,k,i}^\varepsilon {{\mathbf{F}}_n}) + {{\left| {{\beta _{k,i}}} \right|}^2}}  + \delta _k^2}},\forall n,k, \\
			&\qquad{{\mathbf{F}}_n} \succcurlyeq {\mathbf{0}}, {{\mathbf{F}}_n}(l,l) = 1,l = 1,...,(L + 1),\forall n.
	\end{align}}%
\end{subequations}

\noindent As for the objective function (\ref{eq:74a}), the logarithmic term in it can be transformed into
{\footnotesize \begin{align}
		&\underbrace {{{\log }_2}\big(\sum\limits_{i = 1}^K {{\text{Tr}}({\mathbf{\Gamma }}_{n,k,i}^\varepsilon {{\mathbf{F}}_n}) + {{\left| {{\beta _{k,i}}} \right|}^2}} + \delta _k^2\big)}_{{T_{n,k}}({{\mathbf{F}}_n})}\notag \\ 
		&\underbrace { - {{\log }_2}\big(\sum\limits_{i = 1,i \ne k}^K {{\text{Tr}}({\mathbf{\Gamma }}_{n,k,i}^\varepsilon {{\mathbf{F}}_n}) + {{\left| {{\beta _{k,i}}} \right|}^2}} + \delta _k^2\big)}_{{Q_{n,k}}({{\mathbf{F}}_n})},
\end{align}}%
\noindent Then, by utilizing Theorem 1, the lower bound of {\small $ {Q_{n,k}}({{\mathbf{F}}_n}) $} is
{\footnotesize \begin{align}
	Q_{n,k}^{L,(q)}({\mathbf{F}}_n^{(q)}) &= \mathop {\max }\limits_{\eth_{n,k}^{(q)}} - \frac{{\eth_{n,k}^{(q)}\big(\sum\limits_{i = 1,i \ne k}^K {{\text{Tr}}({\mathbf{\Gamma }}_{n,k,i}^\varepsilon {\mathbf{F}}_n^{(q)}) + {{\left| {{\beta _{k,i}}} \right|}^2}} + \delta _k^2\big)}}{{\ln 2}}\notag \\
	& + {\log _2}\eth_{n,k}^{(q)} + (1/\ln 2),\label{eq:76}
\end{align}}%
where 
{\footnotesize \begin{align}
	\eth_{n,k}^{(q)} = {\big(\sum\limits_{i = 1,i \ne k}^K {{\text{Tr}}({\mathbf{\Gamma }}_{n,k,i}^\varepsilon {\mathbf{F}}_n^{(q - 1)}) + {{\left| {{\beta _{k,i}}} \right|}^2}} + \delta _k^2\big)^{ - 1}}.\label{eq:77}
\end{align}}%
\begin{algorithm}[t] 
	\caption{AO-based Algorithm for Solving Problem P13.} 
	\begin{algorithmic}[1]
		\STATE
		\textbf{Initialize}: $ {\mathbf{w}}^{(0)} $, {\small $ ({\mathbf{C}}^{\mathrm{OIA}})^{(0)} $}, $ {\widetilde {\bf f}}_n^{(0)} $, $ \rho_2^{(0)}=0 $, accuracy threshold $ \varepsilon > 0 $, and iteration index $ q = 1 $.
		\STATE
		\textbf{Repeat}
		\STATE
		In the $ q $-th iteration, for given $ \rho_2^{(q-1)} $, {\small $ ({\mathbf{C}}^{\mathrm{OIA}})^{(q-1)} $}, $ {\mathbf{w}}^{(q-1)} $, and $ {\widetilde {\bf f}}_n^{(q-1)} $, solve problem P14 by using Algorithm 2 to obtain $ {\mathbf{v}}_{n,k}^{(q)} $.
		\STATE
		For given $ {\mathbf{v}}_{n,k}^{(q)} $, $ \rho_2^{(q-1)} $, {\small $ {\mathbf{w}}^{(q-1)} $}, and $ {\widetilde {\bf f}}_n^{(q-1)} $, solve problem P17 to obtain {\small $ {\mathbf{W}}^{(q)} $} and {\small $ ({\mathbf{C}}^{\mathrm{OIA}})^{(q)} $}, and decompose $ {\bf{w}}^{(q)} $ from {\small $ {\mathbf{W}}^{(q)} $}.
		\STATE
		For given $ {\mathbf{v}}_{n,k}^{(q)} $, $ {\mathbf{w}}^{(q)} $, {\small $ ({\mathbf{C}}^{\mathrm{OIA}})^{(q)} $}, $ \rho_2^{(q-1)} $, and $ {\widetilde {\bf f}}_n^{(q-1)} $, solve problem P21 to obtain {\small $ {\bf{F}}_n^{(q)} $}, and decompose $ \widetilde{\bf{f}}_n^{(q)} $ from {\small $ {\bf{F}}_n^{(q)} $}.
		\STATE
		{\small $ \rho_2^{(q)}=\frac{B\sum\limits_{n = 1}^N\sum\limits_{k = 1}^K {{\upsilon _k}}{\varpi _{n,k}^{(q)}}R_k^{\mathrm{OIA}}({\mathbf{w}}^{(q)}, ({\mathbf{C}}^{\mathrm{OIA}})^{(q)}, \widetilde{\bf{f}}_n^{(q)})} {{{\cal P}_{\mathrm{total}}^{{\mathrm{OIA}}}}({\bf{w}}^{(q)})} $}.
		\STATE
		Update $ q=q+1 $.
		\STATE
		\textbf{Until} $ \left|\rho_2^{(q)}-\rho_2^{(q-1)}\right| \le \varepsilon $.
	\end{algorithmic} 
	\label{algorithm3}
\end{algorithm}%
Solving problem P21 in the $ q $-th iteration provides a lower bound solution for problem P20.
\begin{subequations}
	{\footnotesize \begin{align}
			&{\text{P}}21:\mathop {\max }\limits_{{\mathbf{F}}_n^{(q)}} \ B\sum\limits_{n = 1}^N \sum\limits_{k = 1}^K {\upsilon _k}{\varpi _{n,k}}({T_{n,k}}({\mathbf{F}}_n^{(q)}) +  Q_{n,k}^{L,(q)}({\mathbf{F}}_n^{(q)})), \\
			&\mathrm{s.t.}\ \frac{{{\text{Tr}}({\mathbf{\Gamma }}_{n,k,k}^\varepsilon {\mathbf{F}}_n^{(q)}) + {{\left| {{\beta _{k,k}}} \right|}^2}}}{{\sum\limits_{i = 1,i \ne k}^K {{\text{Tr}}({\mathbf{\Gamma }}_{n,k,i}^\varepsilon {\mathbf{F}}_n^{(q)}) + {{\left| {{\beta _{k,i}}} \right|}^2}}  + \delta _k^2}} \ge r_k^{\mathrm{OIA}},\forall n,k, \\
			&\quad\quad {\Omega ^{\mathrm{OIA}}} \le \frac{{{\text{Tr}}({\mathbf{\Gamma }}_{n,k,c}^\varepsilon {\mathbf{F}}_n^{(q)}) + {{\left| {{\beta _{k,c}}} \right|}^2}}}{{\sum\limits_{i = 1}^K {{\text{Tr}}({\mathbf{\Gamma }}_{n,k,i}^\varepsilon {\mathbf{F}}_n^{(q)}) + {{\left| {{\beta _{k,i}}} \right|}^2}} + \delta _k^2}},\forall n,k, \\
			&\quad\quad {\mathbf{F}}_n^{(q)} \succcurlyeq {\mathbf{0}}, {\mathbf{F}}_n^{(q)}(l,l) = 1,l = 1,...,L + 1,\forall n.
	\end{align}}%
\end{subequations}

The comprehensive AO-based algorithm devised for solving problem P13 is outlined in Algorithm 3. The convergence of Algorithm 3 can be affirmed by Theorem 3. The detailed proof is omitted here due to space constraints.

\subsection{Computational Complexity}

The computational complexity primarily resides in solving problem P14 using the BnB method and problems P17 and P21 using the SCA method within Algorithm \ref{algorithm3}. For problem P14, the computational complexity of the BnB algorithm shown in Algorithm 2 can be expressed by $ I_B =  \mathcal{O}(NK(\sqrt {{\ell _B}} ({n_B}{\ell _B} + n_B^2{\ell _B} + n_B^3) + NK + 1)) $, where $ {n_B} = NK $ and $ {\ell _B} = NK + N + K $ \cite{6891348,9133120}.

The iteration number required for problem P17 is denoted as $ I_{\mathrm{iter}_3} = \sqrt{(K + 1)M + {\ell _3}} $, where $ {\ell _3} = 2NK + K + 1 $. Within each iteration, the computational complexity is given by $ I_3= \mathcal{O}({n_3}((K + 1){M^3} + {\ell _3}) + n_3^2((K + 1){M^2} + {\ell _3}) + n_3^3) $, where $ {n_3} = (K + 1){M^2} + K $. Therefore, the computational complexity for solving problem P17 is $ \mathcal{O}(I_{\mathrm{iter}_3}I_{3}) $.

For problem P21, the required  iteration number is represented by $ I_{\mathrm{iter}_4} = \sqrt {(L + 1)N + {\ell _4}} $, where $ {\ell _4} = 2NK + L + 1 $. In each iteration, the computational complexity is $ I_4 = \mathcal{O}({n_4}({(L + 1)^3}N + {\ell _4}) + n_4^2({(L + 1)^2}N + {\ell _4}) + n_4^3) $, where $ {n_4} = {(L + 1)^2}N $. Hence, the computational complexity for solving problem P21 is $ {\cal O}(I_{\mathrm{iter}_4}I_{4}) $.

Then, consider $ {I_{\mathrm{AO}_2}} $ as the iteration count of Algorithm 3. The comprehensive computational complexity of Algorithm 3 can be expressed as $ \mathcal{O}({I_{\mathrm{AO}_2}}({I_B} + {I_{\mathrm{iter}_3}}{I_3} + {I_{\mathrm{iter}_4}}{I_4})) $.
\begin{figure}[t]
	\centering	\includegraphics[width=0.92\linewidth,height=0.66\linewidth]{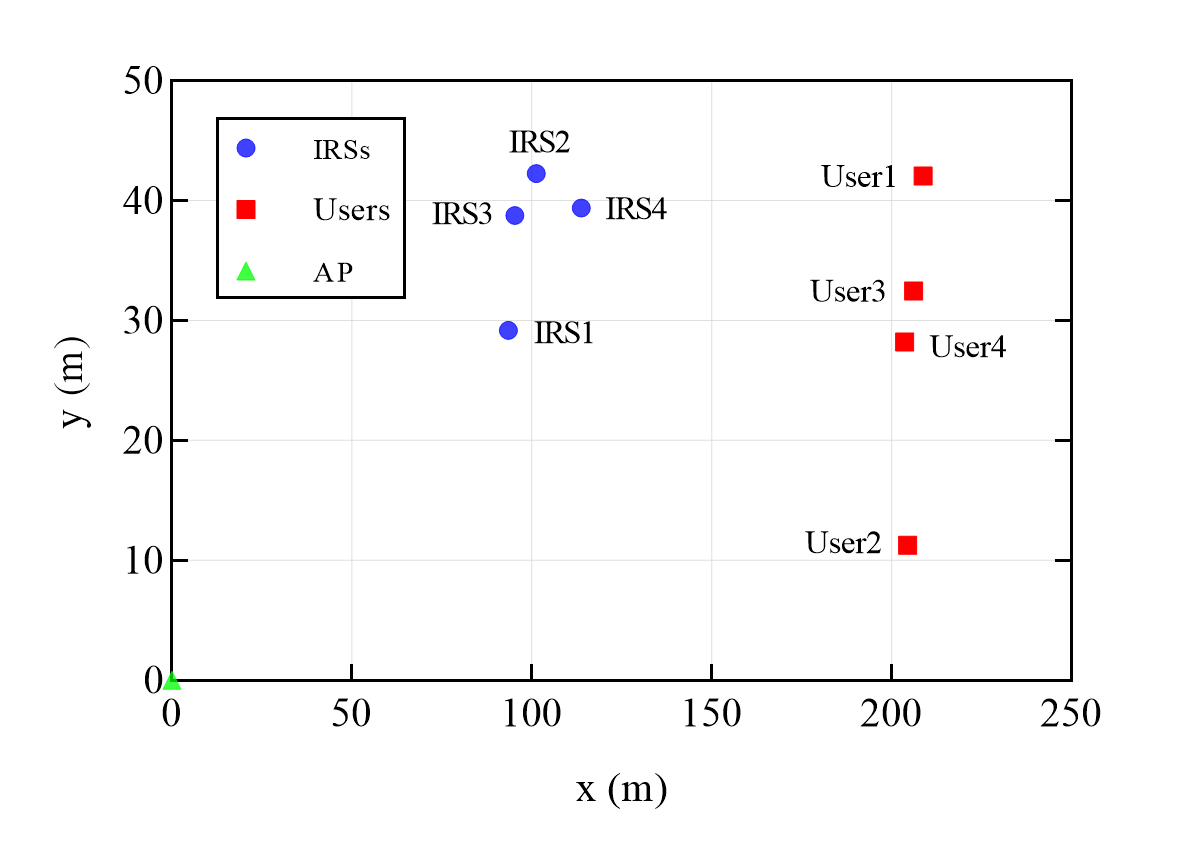}
	\caption{Simulation setup.}
	\label{fig:location}
\end{figure}

\begin{table}[!htbp] \footnotesize
	\centering
	\caption{Simulation parameters.}
	\label{table1}
	\begin{tabular}{cc}
		\toprule
		Parameters& Value\\
		\midrule  %
		Transmission bandwidth, $ B $ $ (\mathrm{kHz}) $ & $ 180 $\\
		Number of transmit antennas, $ M $ & $ 4 $\\
		  Number of reflecting elements of each IRS, $ L $ & $ 30 $\\
		  Total number of IRSs, $ N $ & $ 4 $\\
		  Total number of users, $ K $ & $ 4 $ \\
		The maximum transmit power\\of the AP, $ {P_{\max}} $ $ (\mathrm{dBm}) $ & $ 34 $ \\
            Static power consumption\\of each user, $ P_{\mathrm{user},k} $ $ (\mathrm{dBm}) $ & $ 10 $ \\
		Hardware dissipated power\\at the AP, $ P_{\mathrm{AP}} $ $ (\mathrm{dBm}) $ & $ 37 $ \\
		Per-element static power at the IRS, $ P(b) $ $ (\mathrm{mW}) $ & $ 6 $ \\
		Convergence tolerance & $ 10^{-3} $ \\
		Path-loss for $ {{\bf{g}}_{k}} $ $ (\mathrm{dB}) $ & $ 32.6+36.7\lg\left( d \right) $ \\
		Path-loss for $ {\bf{G}}_n $ and $ {{\bf{h}}_{n,k}} $ $ (\mathrm{dB}) $ & $ 35.6+22\lg\left( d \right) $ \\
		\bottomrule
	\end{tabular}
\end{table}
\section{Numerical Results}

This section presents numerical findings to substantiate the efficacy of the proposed algorithm and to delineate the performance contrast between the EIA and OIA schemes in an RSMA system leveraging multiple IRSs. The four-antenna AP is positioned at coordinates $ (0,0) $ meters and that four single-antenna users are haphazardly distributed within a circular domain concentric at $ (200,30) $ meters with a span of $20$ meters. Furthermore, four distinct IRSs are randomly stationed within a circular expanse, centered at the coordinates $(100,30)$ meters with a range of $20$ meters. Each IRS comprises $30$ reflecting elements. These specifics can be visualized in Fig. \ref{fig:location}. Unless expressly stated otherwise, the remaining critical parameters of the system are summarized in Table \ref{table1}, where $ d $ is the path length. To evaluate the effectiveness of the proposed algorithm, four baseline algorithms are also emulated, namely
\begin{itemize}
	\item Baseline algorithm $ 1 $ (NOMA scheme): Implementation of the NOMA protocol in communication.
	\item Baseline algorithm $ 2 $ (Random transmit beamforming): Random assignment of transmit beamforming while adhering to constraint (\ref{eq:20b}) or (\ref{eq:48b}) \cite{Wei2023}.
	\item Baseline algorithm $ 3 $ (Random phase): Random allocation of phase shifts for the IRSs.
	\item Baseline algorithm $ 4 $ (Without IRS): Absence of IRSs within the system.
\end{itemize}

\begin{figure}[t]
	\centering
	\includegraphics[width=0.92\linewidth,height=0.66\linewidth]{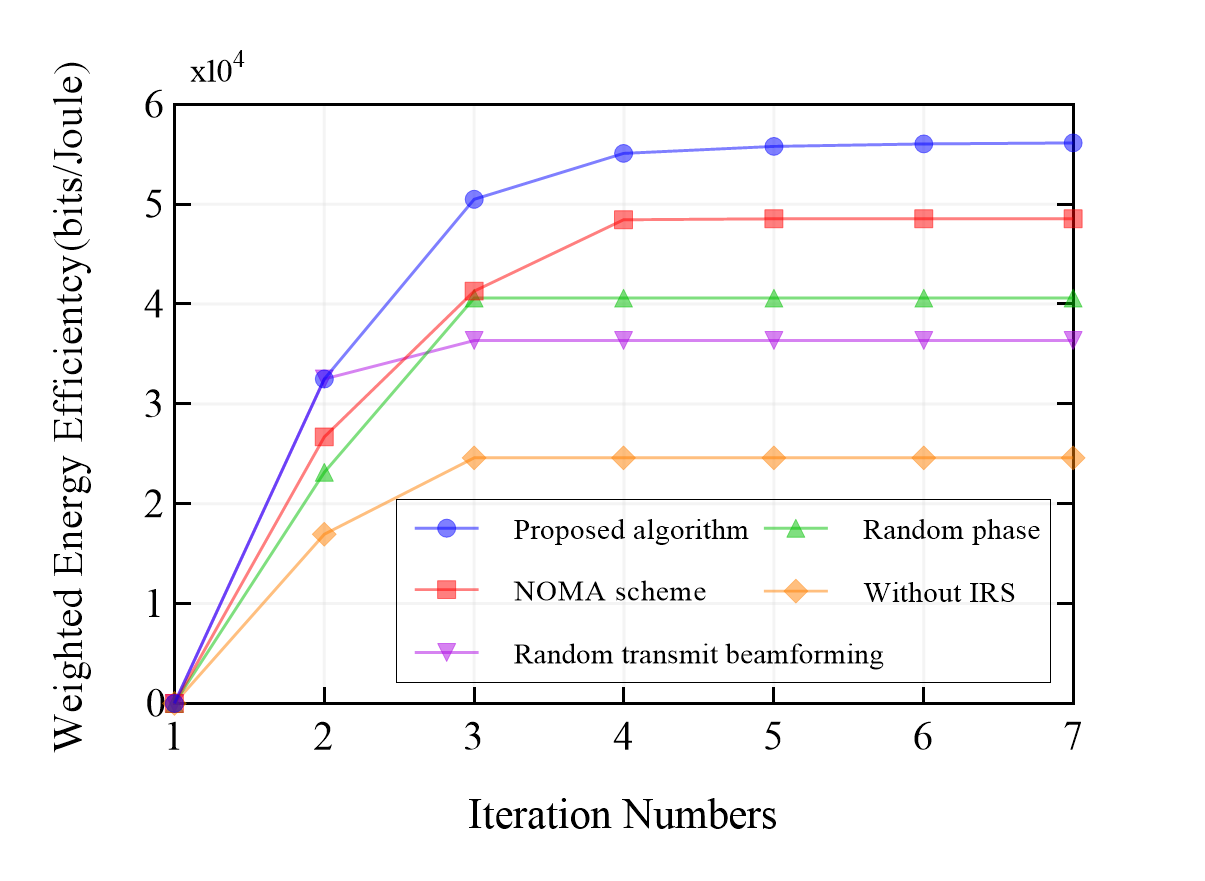}
	\caption{The convergence behaviors of the proposed algorithm and the baseline algorithms in the EIA scheme.}
	\label{fig:eiaconvergence}
\end{figure}
\begin{figure}[t]
	\centering	\includegraphics[width=0.92\linewidth,height=0.66\linewidth]{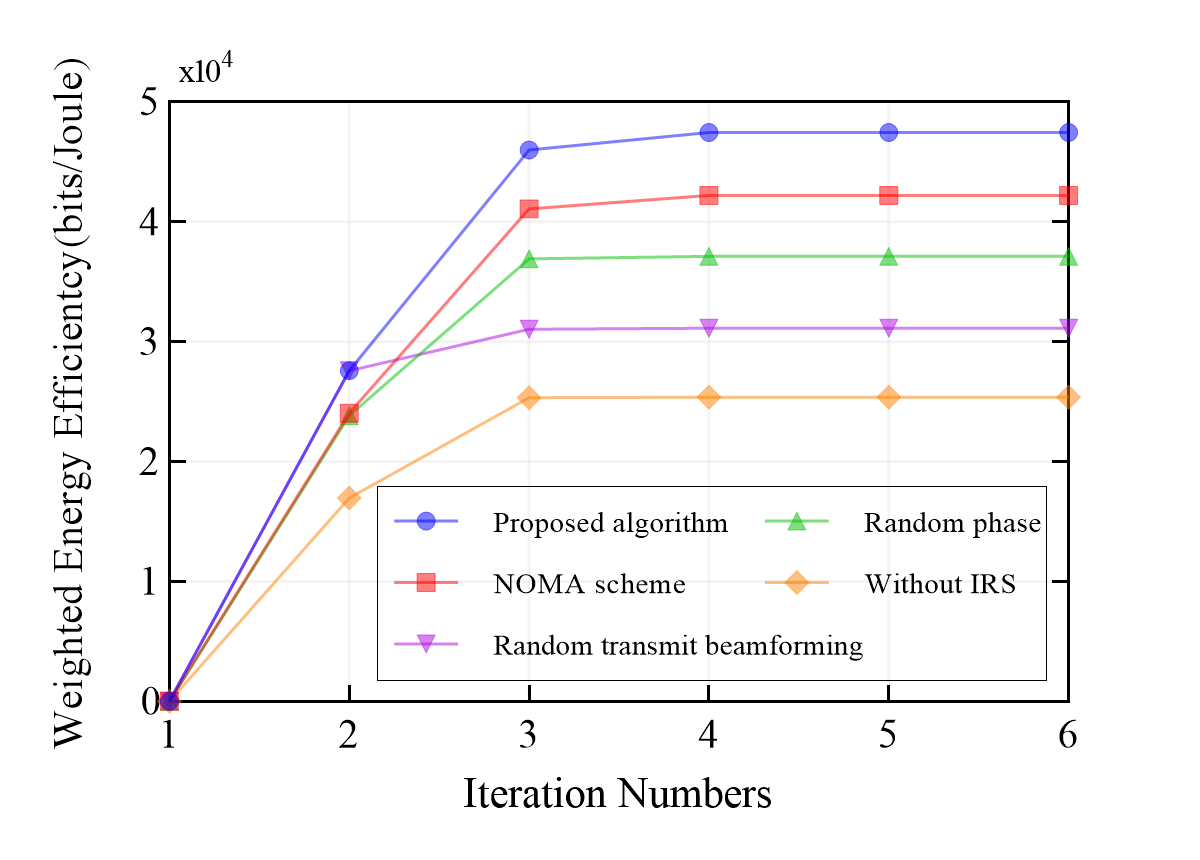}
	\caption{The convergence behaviors of the proposed algorithm and the baseline algorithms in the OIA scheme.}
	\label{fig:oiaconvergence}
\end{figure}

Fig. \ref{fig:eiaconvergence} showcases the convergence dynamics of the proposed algorithm and four baseline algorithms within the EIA scheme when $ L = 30 $ and $ P_{\max} = 34\ \mathrm{dBm} $. It is observed that the weighted EE exhibits an initial increase before converging to a steady state value, and the proposed algorithm achieves quick convergence within seven iterations. This validates the efficacy of the proposed algorithm. It is clear that, in comparison with the algorithm without IRS deployment, the other four algorithms achieve notably superior weighted EE. This enhancement is mainly a result of the IRSs' deployment, which provides additional links to enhance the channel gain from the AP to the users, thereby augmenting the system's weighted EE performance. Furthermore, the achievements in weighted EE by the proposed algorithm and NOMA protocol surpass those of the random phase and random transmit beamforming algorithms. This signifies the crucial need to optimize both the phase shift matrices of the IRSs and the transmit beamforming. In comparison to the NOMA scheme, the proposed algorithm attains remarkable improvements of up to $ 15.645\% $ in terms of weighted EE, due to the utilization of the rate splitting transmission strategy, which offers increased flexibility in resource allocation by employing a superimposed common message stream.

\begin{figure}[t]
	\centering
	\includegraphics[width=0.92\linewidth,height=0.66\linewidth]{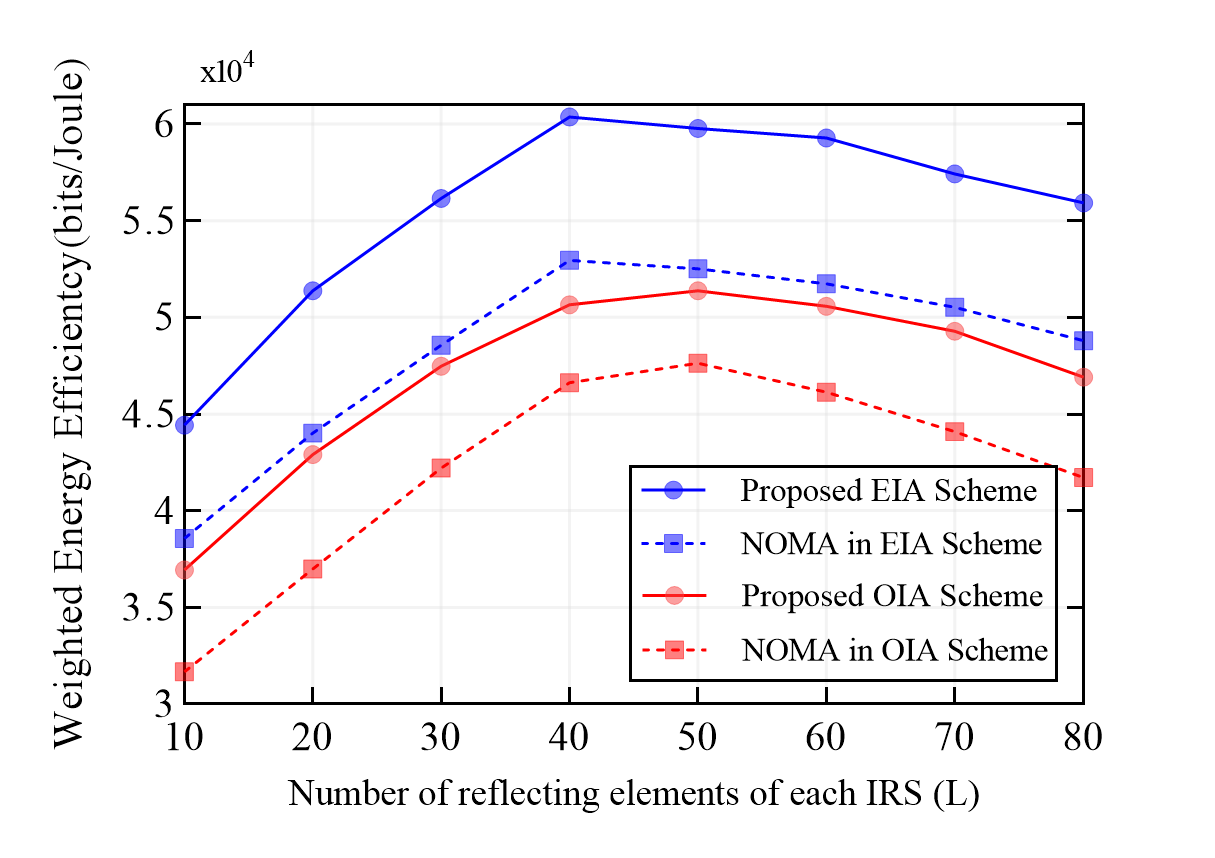}
	\caption{The weighted EE versus the number of reflecting elements of each IRS.}
	\label{fig:relecting-elements}
\end{figure}
\begin{figure}[t]
	\centering
	\includegraphics[width=0.92\linewidth,height=0.66\linewidth]{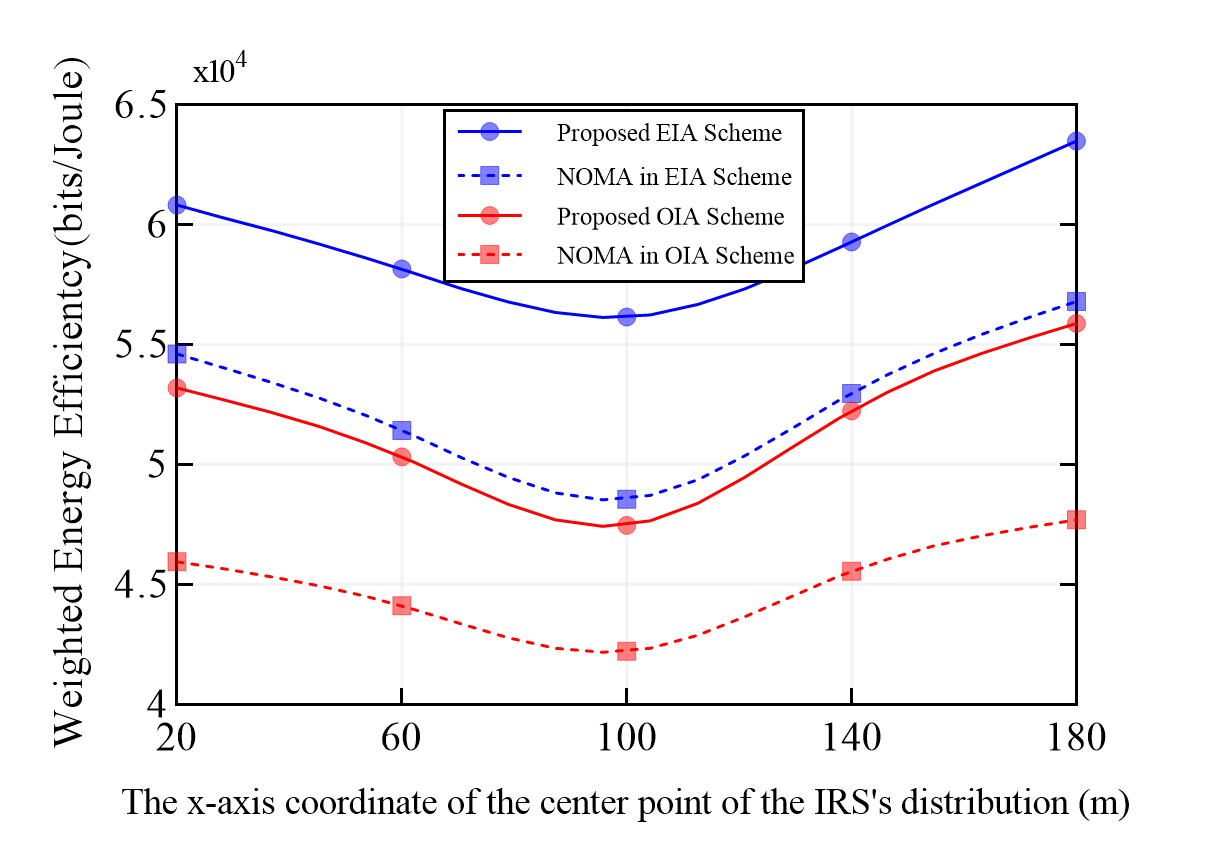}
	\caption{The weighted EE versus the $ x $-coordinate of the central point of the IRSs' deployment range.}
	\label{fig:x-axis}
\end{figure}

Fig. \ref{fig:oiaconvergence} exhibits the convergence behaviors of the proposed algorithm and the baseline algorithms in OIA scheme when $ a = 2 $, $ L = 30 $ and $ P_{\max} = 34\ \mathrm{dBm} $. It is worth noting that the similar conclusions can be drawn as the EIA scheme. A noteworthy distinction lies in the fact that the algorithm proposed in the OIA scheme achieves convergence within six iterations while that of the EIA scheme converges at the $ 7 $-th iteration. Additionally, the performance gap between the proposed algorithm and the NOMA scheme reaches to $ 12.454\% $, further affirming the RSMA's superiority over NOMA.

\begin{figure}[t]
	\centering
	\includegraphics[width=0.92\linewidth,height=0.66\linewidth]{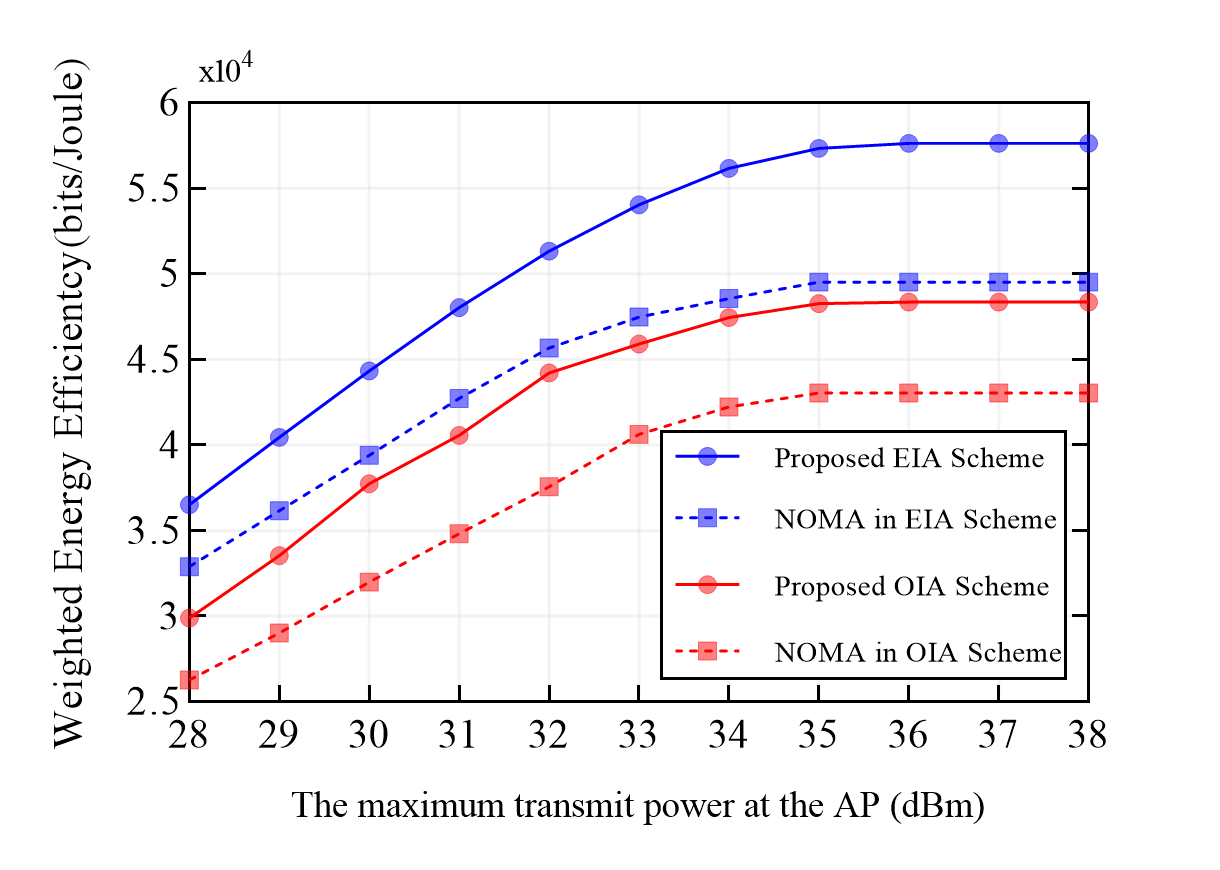}
	\caption{The weighted EE versus the maximum transmit power of AP.}
	\label{fig:transmit-power}
\end{figure}
\begin{figure}[t]
	\centering
	\includegraphics[width=0.92\linewidth,height=0.66\linewidth]{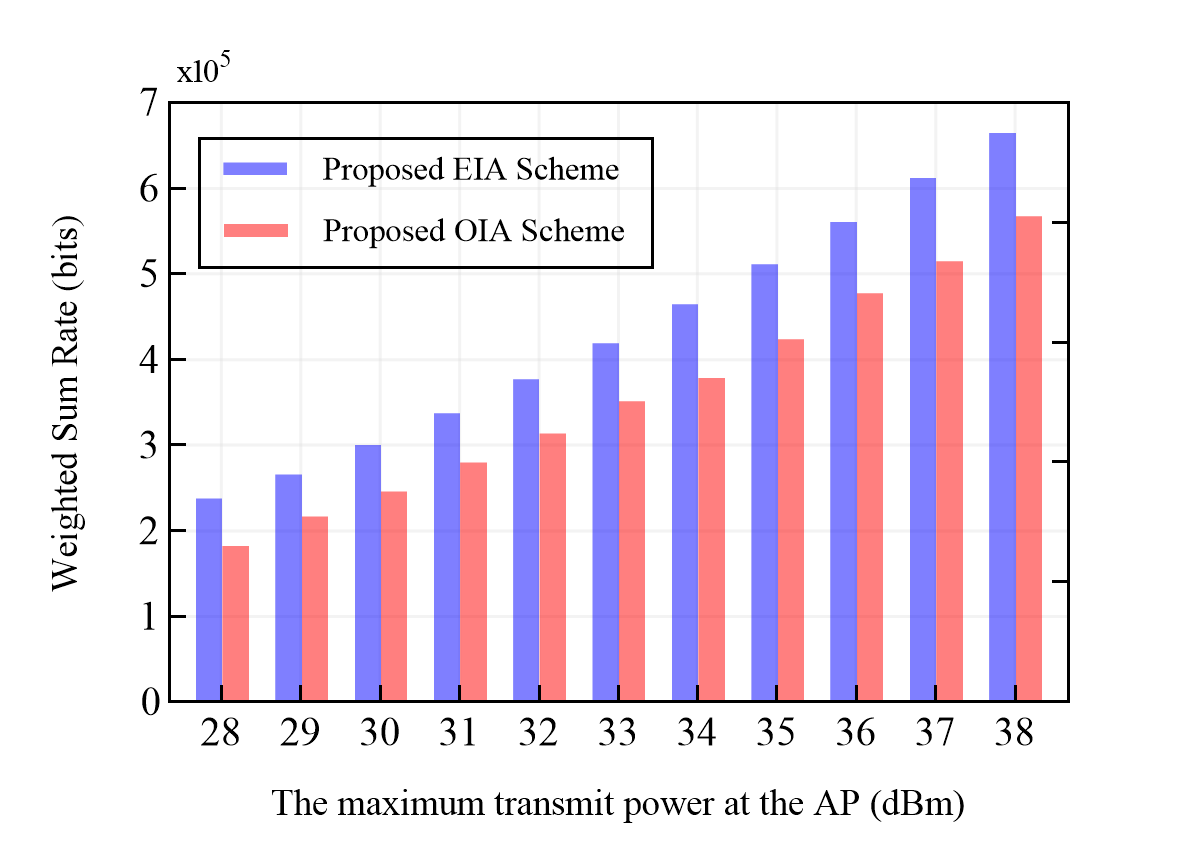}
	\caption{The weighted sum rate versus the maximum transmit power of AP.}
	\label{fig:transmit-power-sum-rate}
\end{figure}

Fig. \ref{fig:relecting-elements} compares the weighted EE as a function of the number of reflecting elements of each IRS for the proposed EIA and OIA schemes when $ a = 2 $ and $ P_{\max} = 34\ \mathrm{dBm} $. Observations indicate that as the quantity of reflecting elements on each IRS increases, the weighted EE of both schemes show an upward-then-downward trend. The upward trend can be explained as that when each IRS possess a fewer number of reflecting elements, the energy consumed by the IRSs is very low compared to the static power consumption in the system. Meanwhile, the IRSs with more reflecting elements can provide better transmission rate, thereby improving the system weighted EE. In this process, the EIA scheme consistently outperforms the OIA scheme. However, once the number of reflecting elements exceeds $ 40 $, the weighted EE in the EIA scheme exhibits a downward trend, and this trend also occurs when the number of reflecting elements of IRSs exceeds $ 50 $ in the OIA scheme. The reason is that while the achievable rate climbs with the quantity of reflecting elements, the energy consumption from the additional elements outweighs the associated transmission rate gain, leading to a gradual decrease in the system weighted EE.
The EE of RSMA-based schemes achieve better performance compared with their corresponding NOMA-based schemes. The reason is that the more customized rate allocation optimization in RSMA may improve the data rate, reduce the interference, and thus achieve better communication performance compared with NOMA-based schemes. The curves in the following figures also show the same observation.

Fig. \ref{fig:x-axis} exhibits the weighted EE versus the $ x $-axis coordinate of the center point of IRSs when $ a = 2 $, $ L = 30 $, and $ P_{\max} = 34\ \mathrm{dBm} $. As shown in Fig. \ref{fig:location}, all IRSs are randomly deployed within a circular area centered at a fixed point with radius $ 20\ \mathrm{m} $, and they are located between the AP and users. If the $ x $-axis coordinate of the center point is smaller, the IRSs are much closer to the AP, otherwise, they are near to the users. Contrary to the notion that the optimal performance is achieved when the relay is positioned in the middle of the targeted transceiver link, our findings suggest that distributed IRSs positioned nearer to either the AP or users exhibit superior EE performance, as depicted in Fig. \ref{fig:x-axis}. Moreover, it is evident that user-side deployment (where IRSs are in proximity to users) outperforms AP-side deployment (where IRSs are closer to the AP).

Fig. \ref{fig:transmit-power} depicts the weighted EE versus the maximum transmit power of the AP in the proposed EIA and OIA schemes with $ a = 2 $ and $ L = 30 $. As expected, the EIA scheme achieves much higher weighted EE than the OIA scheme. This is because there are four IRSs to help each transmission between the AP and users in the EIA scheme, however, only one IRS is helped to one transmission in the OIA scheme. In addition, from Fig. \ref{fig:relecting-elements} we know that the total throughput obtained by the aid of IRSs with $ 30 $ reflecting elements dominates the energy consumed by them. Therefore, the EIA scheme outperforms the OIA scheme in this situation. In Fig. \ref{fig:transmit-power}, it is observed that the weighted EE of both the EIA and OIA schemes initially increases with the transmit power budget $ P_{\max}$, but as $ P_{\max} $ is raised to $ 35\ \mathrm{dBm} $, both schemes exhibit nearly saturated weighted EE. This behavior can be attributed to the following reasoning: when $ P_{\max} $ is comparatively low, an increase in $ P_{\max} $ leads to a boost in the received signal power at the users, which in turn improves their signal-to-noise ratio. And because $ P_{\max} $ is low, the power consumption brought by it is also small, thereby facilitating an enhancement in weighted EE. However, when the value of $ P_{\max} $ increases to a certain value, the benefits conferred by increasing $ P_{\max} $ are counterbalanced by the corresponding increase in power consumption. At this time, the system's EE no longer experiences growth but instead reaches a state of saturation.

As an additional proof to Fig. \ref{fig:transmit-power}, the impact of AP transmit power on system weighted sum rate is shown in Fig. \ref{fig:transmit-power-sum-rate} when $ a = 2 $ and $ L = 30 $. It can be observed that as the AP transmit power increases, the achievable weighted sum rate also experiences a gradual escalation under both schemes. This phenomenon arises due to the larger transmit power, leading to stronger received signals and consequently yielding higher achievable weighted sum rate for the system. However, the system power consumption also increases due to the higher power required to generate the signal at higher levels. Thus, this will lead to a trade-off between the achievable weighted sum rate and system power consumption, as illustrated in Fig. \ref{fig:transmit-power}.

\section {Conclusion}

This paper investigates the weighted system EE maximization problem under both the EIA and OIA schemes, respectively, in a multi-IRS assisted RSMA system. The optimization process in both schemes involves the joint optimization of transmit beamforming at the AP, the phase shift matrices at the IRSs, and the common achievable rate at the users. And the IRS association is additionally optimized in OIA scheme. To tackle the formulated non-convex problems, the Dinkelbach, BnB, SCA and SDR methods are introduced. And the efficient algorithms based on AO framework in both schemes are proposed. Numerical results validate the proposed algorithms achieves significant performance improvement compared with the baseline algorithms, and the EIA scheme is proved outperforms OIA scheme under various circumstances. Furthermore, based on the numeriacal results, we can inferred that the quantity of reflecting elements in each IRS should not exceed a certain limit (e.g., $ L \le 40 $ in EIA scheme in this paper). Additionally, it is recommended to maintain the maximal transmit power at the AP within a reasonable range (e.g., $ P_{\max} \le 35\ \mathrm{dBm} $ in this paper). Lastly, the deployment position of IRSs should be in close proximity to either the AP or the users.

\begin{appendices}
	\section{Proof of Theorem \ref{Theorem:2}}
	\begin{Proof}
		As an illustration, we will demonstrate the proof of Theorem 2 with $ {\bf W}_c^* $. Following a comparable methodology as outlined in {\rm{\cite{9195771,9133130}}}, and eliminating extraneous and irrelevant terms, the Lagrangian function of problem {\rm{P5}} with respect to $ {\bf W}_c^* $ can be formulated as follows
		\begin{align}
			L({\mathbf{W}}_c^ \star ) & = (\lambda _1 + {\rho _1}){\text{Tr}}({\mathbf{W}}_c^ \star ) - \sum\limits_{k = 1}^K {{\lambda _{2,k}}{\text{Tr(}}{{\mathbf{B}}_k}{\mathbf{W}}_c^ \star )} \notag \\ 
			& - {\text{Tr}}({{\mathbf{Q}}_c}{\mathbf{W}}_c^ \star ),\label{eq:79}
		\end{align}
		where $ {\lambda _1} \ge 0 $ and $ {\lambda _{2,k}} \ge 0 $ represent the dual variables linked to constraints {\rm{(\ref{eq:37b})}} and {\rm{(\ref{eq:37d})}}, correspondingly. Additionally, $ {\mathbf{Q}}_c \succcurlyeq {\mathbf{0}} $ signifies the dual variable matrix associated with constraint {\rm{(\ref{eq:37f})}}. Our attention now shifts to the Karush-Kuhn-Tucker (KKT) conditions, imperative for the proof.
		\begin{subequations}
			\begin{align}
				&K1:{\lambda _1^\star} \ge 0,{\lambda _{2,k}^\star} \ge 0, {\mathbf{Q}}_c^ \star  \succcurlyeq {\mathbf{0}}, \\
				&K2:{\mathbf{Q}}_c {\mathbf{W}}_c^ \star = {\mathbf{0}}, \\
				&K3:{\nabla _{{\mathbf{W}}_c^ \star }}L({\mathbf{W}}_c^ \star ) = 0.
			\end{align}
		\end{subequations}
		
		The KKT condition $ K3 $ can be rewritten as
		{\small \begin{align}
				{\nabla _{{\mathbf{W}}_c^ \star }}L({\mathbf{W}}_c^ \star ) = ({\lambda _1} + {\rho _1}{\text{)}}{\mathbf{I}}_M - \sum\limits_{k = 1}^K {{\lambda _{2,k}}{{\mathbf{B}}_k}} - {\mathbf{Q}}_c^ \star = 0,
		\end{align}}
		
		Denote $ \eta = {\lambda _1} + {\rho _1} $ and {\small $ {\bf{T}} = \sum\limits_{k = 1}^K {{\lambda _{2,k}}{{\mathbf{B}}_k}} $, then $ {\mathbf{Q}}_c^ \star $} can be expressed by
		{\small \begin{align}
				{\mathbf{Q}}_c^ \star = \eta{\mathbf{I}}_M - \bf{T}.
		\end{align}}
		
		From the KKT condition $K1$, it is determined that ${\mathbf{Q}}_c^*$ must be positive semi-definite, and $\eta \ge \widetilde \lambda_{\bf T}^{\rm max} \ge 0$, where $\widetilde \lambda_{\bf T}^{\rm max}$ represents the maximum eigenvalue of ${\bf T}$. According to the KKT condition $K2$, the columns of ${\mathbf{W}}_c^*$ must lie in the null space of ${\mathbf{Q}}_c^*$. If $\eta > \widetilde \lambda_{\bf T}^{\rm max}$, ${\mathbf{Q}}_c^*$ will maintain a full rank. However, this scenario would result in the solution ${\mathbf{W}}_c^* = {\mathbf{0}}$. Evidently, ${\mathbf{W}}_c^* = {\mathbf{0}}$ would not leverage the benefits of RSMA, and thus it is not the desired solution. Conversely, if $\eta = \widetilde \lambda_{\bf T}^{\rm max}$, this would lead to ${\rm Rank}({\mathbf{Q}}_c^*) = M-1$, causing the rank of the matrix ${\mathbf{W}}_c^*$ equals to $1$. By employing a similar approach, it can be readily demonstrated that $\{ {\mathbf{W}}_k^*, \forall k \}$ satisfies Theorem 2. Due to space constraints, the detailed proof process is omitted.
		
		The proof is completed.
		$\hfill \square$
	\end{Proof}
	
	\section{Proof of Theorem \ref{Theorem:3}}
	\begin{Proof}
		In order to prove Theorem 3, a similar methodology as demonstrated in {\rm{\cite{8811733,Kan2023}}} is introduced.
		Define {\small $ \mathrm{EE}({\bf W}^{(t)},({\bf C}^{\mathrm{EIA}})^{(t)},{\bf f}^{(t)}) $}, {\small $ {\rm{g}}_{\mathrm{P4}}({\bf W}^{(t)},({\bf C}^{\mathrm{EIA}})^{(t)}) $} and {\small $ {\rm{g}}_{\mathrm{P9}}({\bf f}^{(t)}) $} as the objective function of problem {\rm{P1}}, {\rm{P4}} and {\rm{P9}} in the $ t $-th iteration, respectively. Therefore, the following expression always holds
		\begin{equation}
			{\small \begin{split}
					\mathrm{EE}&({\bf W}^{(t+1)},({\bf C}^{\mathrm{EIA}})^{(t+1)},{\bf f}^{(t+1)})\\
					&\mathop = \limits^{(a)}{\rm{g}}_{\mathrm{P9}}({\bf f}^{(t+1)})/{({\cal P}_{\mathrm{total}}^{\mathrm{EIA}})^{(t+1)}}\\ 
					&\mathop \ge \limits^{(b)} {\rm{g}}_{\mathrm{P9}}({\bf f}^{(t)})/{({\cal P}_{\mathrm{total}}^{\mathrm{EIA}})^{(t+1)}}\\ 
					&\mathop = \limits^{(c)} \mathrm{EE}({\bf W}^{(t+1)},({\bf C}^{\mathrm{EIA}})^{(t+1)},{\bf f}^{(t)}), \label{eq:84}
			\end{split}}%
		\end{equation}
		where $ (a) $ and $ (c) $ follow the fact that with given {\small $ ({\bf W},{\bf C}^{\mathrm{EIA}}) $} problems P1 and P9 are equivalent; and $ (b) $ holds since $ {\bf f}^{(t+1)} $ is the optimized solution. Similarly, with given $ {\bf f}^{(t)} $, we have
		\begin{equation}
			{\small \begin{split}
					\mathrm{EE}&({\bf W}^{(t+1)},({\bf C}^{\mathrm{EIA}})^{(t+1)},{\bf f}^{(t)})\\
					& = {\rm{g}}_{\mathrm{P4}}({\bf W}^{(t+1)},({\bf C}^{\mathrm{EIA}})^{(t+1)})\\
					& \ge {\rm{g}}_{\mathrm{P4}}({\bf W}^{(t)},({\bf C}^{\mathrm{EIA}})^{(t)})\\
					& = \mathrm{EE}({\bf W}^{(t)},({\bf C}^{\mathrm{EIA}})^{(t)},{\bf f}^{(t)}). \label{eq:85}
			\end{split}}
		\end{equation}
		
		From inequalities {\rm{(\ref{eq:84})}} and {\rm{(\ref{eq:85})}}, it is observed that the value of the objective function for problem P1 increases monotonically as the number of iterations progresses. Given that the system's resources are limited, it follows that Algorithm 1 will ultimately reach a stationary point.
		
		The proof is completed.
		$\hfill \square$
	\end{Proof}
\end{appendices}

\bibliographystyle{IEEEtran}
\bibliography{reference}

\begin{thebibliography}{10}
\providecommand{\url}[1]{#1}
\csname url@samestyle\endcsname
\providecommand{\newblock}{\relax}
\providecommand{\bibinfo}[2]{#2}
\providecommand{\BIBentrySTDinterwordspacing}{\spaceskip=0pt\relax}
\providecommand{\BIBentryALTinterwordstretchfactor}{4}
\providecommand{\BIBentryALTinterwordspacing}{\spaceskip=\fontdimen2\font plus
\BIBentryALTinterwordstretchfactor\fontdimen3\font minus
  \fontdimen4\font\relax}
\providecommand{\BIBforeignlanguage}[2]{{%
\expandafter\ifx\csname l@#1\endcsname\relax
\typeout{** WARNING: IEEEtran.bst: No hyphenation pattern has been}%
\typeout{** loaded for the language `#1'. Using the pattern for}%
\typeout{** the default language instead.}%
\else
\language=\csname l@#1\endcsname
\fi
#2}}
\providecommand{\BIBdecl}{\relax}
\BIBdecl

\bibitem{8808168}
K.~B. Letaief, W.~Chen, Y.~Shi, J.~Zhang, and Y.-J.~A. Zhang, ``The roadmap to
  {6G}: {AI} empowered wireless networks,'' \emph{IEEE Communications
  Magazine}, vol.~57, no.~8, pp. 84--90, 2019.

\bibitem{9246254}
H.~Xie, J.~Xu, and Y.-F. Liu, ``Max-min fairness in {IRS}-aided multi-cell
  {MISO} systems with joint transmit and reflective beamforming,'' \emph{IEEE
  Transactions on Wireless Communications}, vol.~20, no.~2, pp. 1379--1393,
  2021.

\bibitem{9644606}
Y.~Zou, Y.~Long, S.~Gong, D.~T. Hoang, W.~Liu, W.~Cheng, and D.~Niyato,
  ``Robust beamforming optimization for self-sustainable intelligent reflecting
  surface assisted wireless networks,'' \emph{IEEE Transactions on Cognitive
  Communications and Networking}, vol.~8, no.~2, pp. 856--870, 2022.

\bibitem{8537962}
F.~Wang, J.~Xu, and Z.~Ding, ``Multi-antenna {NOMA} for computation offloading
  in multiuser mobile edge computing systems,'' \emph{IEEE Transactions on
  Communications}, vol.~67, no.~3, pp. 2450--2463, 2019.

\bibitem{9918632}
Z.~Wang, L.~Liu, S.~Zhang, and S.~Cui, ``Massive {MIMO} communication with
  intelligent reflecting surface,'' \emph{IEEE Transactions on Wireless
  Communications}, vol.~22, no.~4, pp. 2566--2582, 2023.

\bibitem{9154573}
J.~Zhao, M.~M. Amiri, and D.~Gündüz, ``Multi-antenna coded content delivery
  with caching: A low-complexity solution,'' \emph{IEEE Transactions on
  Wireless Communications}, vol.~19, no.~11, pp. 7484--7497, 2020.

\bibitem{artirsma}
Y.~Mao, B.~Clerckx, and V.~Li, ``Rate-splitting multiple access for downlink
  communication systems: Bridging, generalizing and outperforming {SDMA} and
  {NOMA},'' \emph{EURASIP Journal on Wireless Communications and Networking},
  vol. 2018, 2018.

\bibitem{als20216}
M.~Alsabah, M.~A. Naser, B.~M. Mahmmod, S.~H. Abdulhussain, M.~R. Eissa,
  A.~Al-Baidhani, N.~K. Noordin, S.~M. Sait, K.~A. Al-Utaibi, and F.~Hashim,
  ``{6G} wireless communications networks: A comprehensive survey,'' \emph{IEEE
  Access}, vol.~9, pp. 148\,191--148\,243, 2021.

\bibitem{9831440}
Y.~Mao, O.~Dizdar, B.~Clerckx, R.~Schober, P.~Popovski, and H.~V. Poor,
  ``Rate-splitting multiple access: Fundamentals, survey, and future research
  trends,'' \emph{IEEE Communications Surveys \& Tutorials}, vol.~24, no.~4,
  pp. 2073--2126, 2022.

\bibitem{9832611}
A.~Mishra, Y.~Mao, O.~Dizdar, and B.~Clerckx, ``Rate-splitting multiple access
  for {6G}—part {I}: Principles, applications and future works,'' \emph{IEEE
  Communications Letters}, vol.~26, no.~10, pp. 2232--2236, 2022.

\bibitem{9226406}
S.~Naser, P.~C. Sofotasios, L.~Bariah, W.~Jaafar, S.~Muhaidat, M.~Al-Qutayri,
  and O.~A. Dobre, ``Rate-splitting multiple access: Unifying {NOMA} and {SDMA}
  in {MISO} {VLC} channels,'' \emph{IEEE Open Journal of Vehicular Technology},
  vol.~1, pp. 393--413, 2020.

\bibitem{9663192}
A.~Mishra, Y.~Mao, O.~Dizdar, and B.~Clerckx, ``Rate-splitting multiple access
  for downlink multiuser {MIMO}: Precoder optimization and phy-layer design,''
  \emph{IEEE Transactions on Communications}, vol.~70, no.~2, pp. 874--890,
  2022.

\bibitem{9519632}
H.~Xie, J.~Xu, Y.-F. Liu, L.~Liu, and D.~W.~K. Ng, ``User grouping and
  reflective beamforming for {IRS}-aided {URLLC},'' \emph{IEEE Wireless
  Communications Letters}, vol.~10, no.~11, pp. 2533--2537, 2021.

\bibitem{10440056}
X.~Song, X.~Qin, J.~Xu, and R.~Zhang, ``Cramér-rao bound minimization for
  {IRS}-enabled multiuser integrated sensing and communications,'' \emph{IEEE
  Transactions on Wireless Communications}, pp. 1--1, 2024.

\bibitem{8910627}
Q.~Wu and R.~Zhang, ``Towards smart and reconfigurable environment: Intelligent
  reflecting surface aided wireless network,'' \emph{IEEE Communications
  Magazine}, vol.~58, no.~1, pp. 106--112, 2020.

\bibitem{9907933}
A.~Gashtasbi, M.~M. Silva, and R.~Dinis, ``An overview of intelligent
  reflecting surfaces for future wireless systems,'' in \emph{2022 13th
  International Symposium on Communication Systems, Networks and Digital Signal
  Processing (CSNDSP)}, 2022, pp. 314--319.

\bibitem{9326394}
Q.~Wu, S.~Zhang, B.~Zheng, C.~You, and R.~Zhang, ``Intelligent reflecting
  surface-aided wireless communications: A tutorial,'' \emph{IEEE Transactions
  on Communications}, vol.~69, no.~5, pp. 3313--3351, 2021.

\bibitem{9095301}
A.-A.~A. Boulogeorgos and A.~Alexiou, ``Performance analysis of reconfigurable
  intelligent surface-assisted wireless systems and comparison with relaying,''
  \emph{IEEE Access}, vol.~8, pp. 94\,463--94\,483, 2020.

\bibitem{9623452}
Y.~Zhao, B.~Clerckx, and Z.~Feng, ``{IRS}-aided {SWIPT}: Joint waveform, active
  and passive beamforming design under nonlinear harvester model,'' \emph{IEEE
  Transactions on Communications}, vol.~70, no.~2, pp. 1345--1359, 2022.

\bibitem{9751048}
M.~Shehab, B.~S. Ciftler, T.~Khattab, M.~M. Abdallah, and D.~Trinchero, ``Deep
  reinforcement learning powered {IRS}-assisted downlink {NOMA},'' \emph{IEEE
  Open Journal of the Communications Society}, vol.~3, pp. 729--739, 2022.

\bibitem{10288203}
Y.~Shen, C.~Wang, W.~Zang, L.~Xue, B.~Yang, and X.~Guan, ``Outage constrained
  max-min secrecy rate optimization for {IRS}-aided {SWIPT} systems with
  artificial noise,'' \emph{IEEE Internet of Things Journal}, pp. 1--1, 2023.

\bibitem{9417469}
X.~Zhang, Y.~Shen, B.~Yang, W.~Zang, and S.~Wang, ``{DRL} based data offloading
  for intelligent reflecting surface aided mobile edge computing,'' in
  \emph{2021 IEEE Wireless Communications and Networking Conference (WCNC)},
  2021, pp. 1--7.

\bibitem{10497119}
Y.~Fang, S.~Zhang, X.~Li, X.~Yu, J.~Xu, and S.~Cui, ``Multi-{IRS}-enabled
  integrated sensing and communications,'' \emph{IEEE Transactions on
  Communications}, pp. 1--1, 2024.

\bibitem{9832618}
H.~Li, Y.~Mao, O.~Dizdar, and B.~Clerckx, ``Rate-splitting multiple access for
  {6G}—part {III}: Interplay with reconfigurable intelligent surfaces,''
  \emph{IEEE Communications Letters}, vol.~26, no.~10, pp. 2242--2246, 2022.

\bibitem{9849099}
M.~R. Camana, C.~E. Garcia, and I.~Koo, ``Rate-splitting multiple access in a
  {MISO SWIPT} system assisted by an intelligent reflecting surface,''
  \emph{IEEE Transactions on Green Communications and Networking}, vol.~6,
  no.~4, pp. 2084--2099, 2022.

\bibitem{10000454}
B.~K.~S. Lima, R.~Dinis, D.~B. Da~Costa, M.~Beko, R.~Oliveira, R.~Vigelis, and
  M.~Debbah, ``Rate-splitting multiple access networks assisted by aerial
  intelligent reflecting surfaces,'' in \emph{2022 IEEE Latin-American
  Conference on Communications (LATINCOM)}, 2022, pp. 1--6.

\bibitem{9760044}
A.~Bansal, N.~Agrawal, and K.~Singh, ``Rate-splitting multiple access for
  {UAV}-based {RIS}-enabled interference-limited vehicular communication
  system,'' \emph{IEEE Transactions on Intelligent Vehicles}, vol.~8, no.~1,
  pp. 936--948, 2023.

\bibitem{10008582}
M.~Katwe, K.~Singh, B.~Clerckx, and C.-P. Li, ``Rate splitting multiple access
  for energy efficient {RIS}-aided multi-user short-packet communications,'' in
  \emph{2022 IEEE Globecom Workshops (GC Wkshps)}, 2022, pp. 644--649.

\bibitem{10032267}
R.~Zhang, K.~Xiong, Y.~Lu, P.~Fan, D.~W.~K. Ng, and K.~B. Letaief, ``Energy
  efficiency maximization in {RIS}-assisted {SWIPT} networks with {RSMA}: A
  {PPO}-based approach,'' \emph{IEEE Journal on Selected Areas in
  Communications}, vol.~41, no.~5, pp. 1413--1430, 2023.

\bibitem{9145189}
Z.~Yang, J.~Shi, Z.~Li, M.~Chen, W.~Xu, and M.~Shikh-Bahaei, ``Energy efficient
  rate splitting multiple access ({RSMA}) with reconfigurable intelligent
  surface,'' in \emph{2020 IEEE International Conference on Communications
  Workshops (ICC Workshops)}, 2020, pp. 1--6.

\bibitem{8811733}
Q.~Wu and R.~Zhang, ``Intelligent reflecting surface enhanced wireless network
  via joint active and passive beamforming,'' \emph{IEEE Transactions on
  Wireless Communications}, vol.~18, no.~11, pp. 5394--5409, 2019.

\bibitem{huang2019reconfigurable}
C.~Huang, A.~Zappone, G.~C. Alexandropoulos, M.~Debbah, and C.~Yuen,
  ``Reconfigurable intelligent surfaces for energy efficiency in wireless
  communication,'' \emph{IEEE Transactions on Wireless Communications},
  vol.~18, no.~8, pp. 4157--4170, 2019.

\bibitem{9133120}
J.~Liu, K.~Xiong, Y.~Lu, D.~W.~K. Ng, Z.~Zhong, and Z.~Han, ``Energy efficiency
  in secure {IRS}-aided {SWIPT},'' \emph{IEEE Wireless Communications Letters},
  vol.~9, no.~11, pp. 1884--1888, 2020.

\bibitem{9882159}
F.~Guo, C.~Huang, Y.~Xu, Z.~Yang, and M.~Shikh-Bahaei, ``Secure beamforming
  design for reconfigurable intelligent surface assisted downlink
  transmission,'' in \emph{2022 IEEE International Conference on Communications
  Workshops (ICC Workshops)}, 2022, pp. 1--6.

\bibitem{9195771}
H.~Fu, S.~Feng, W.~Tang, and D.~W.~K. Ng, ``Robust secure beamforming design
  for two-user downlink {MISO} rate-splitting systems,'' \emph{IEEE
  Transactions on Wireless Communications}, vol.~19, no.~12, pp. 8351--8365,
  2020.

\bibitem{7015632}
R.~Feng, Q.~Li, Q.~Zhang, and J.~Qin, ``Robust secure beamforming in {MISO}
  full-duplex two-way secure communications,'' \emph{IEEE Transactions on
  Vehicular Technology}, vol.~65, no.~1, pp. 408--414, 2016.

\bibitem{6891348}
K.-Y. Wang, A.~M.-C. So, T.-H. Chang, W.-K. Ma, and C.-Y. Chi, ``Outage
  constrained robust transmit optimization for multiuser {MISO} downlinks:
  Tractable approximations by conic optimization,'' \emph{IEEE Transactions on
  Signal Processing}, vol.~62, no.~21, pp. 5690--5705, 2014.

\bibitem{boyd2007branch}
S.~Boyd and J.~Mattingley, ``Branch and bound methods,'' \emph{Notes for
  EE364b, Stanford University}, vol. 2006, p.~07, 2007.

\bibitem{7934461}
Z.~Wei, D.~W.~K. Ng, J.~Yuan, and H.-M. Wang, ``Optimal resource allocation for
  power-efficient {MC-NOMA} with imperfect channel state information,''
  \emph{IEEE Transactions on Communications}, vol.~65, no.~9, pp. 3944--3961,
  2017.

\bibitem{5200968}
Y.-B. Lin, T.-H. Chiu, and Y.~T. Su, ``Optimal and near-optimal resource
  allocation algorithms for {OFDMA} networks,'' \emph{IEEE Transactions on
  Wireless Communications}, vol.~8, no.~8, pp. 4066--4077, 2009.

\bibitem{Xu2022}
D.~Xu, V.~Jamali, X.~Yu, D.~W.~K. Ng, and R.~Schober, ``Optimal resource
  allocation design for large {IRS}-assisted {SWIPT} systems: A scalable
  optimization framework,'' \emph{IEEE Transactions on Communications},
  vol.~70, no.~2, pp. 1423--1441, 2022.

\bibitem{Wei2023}
W.~Wei, X.~Pang, J.~Tang, N.~Zhao, X.~Wang, and A.~Nallanathan, ``Secure
  transmission design for aerial {IRS} assisted wireless networks,'' \emph{IEEE
  Transactions on Communications}, vol.~71, no.~6, pp. 3528--3540, 2023.

\bibitem{9133130}
X.~Yu, D.~Xu, Y.~Sun, D.~W.~K. Ng, and R.~Schober, ``Robust and secure wireless
  communications via intelligent reflecting surfaces,'' \emph{IEEE Journal on
  Selected Areas in Communications}, vol.~38, no.~11, pp. 2637--2652, 2020.

\bibitem{Kan2023}
T.-Y. Kan, R.~Y. Chang, F.-T. Chien, B.-J. Chen, and H.~V. Poor, ``Hybrid relay
  and reconfigurable intelligent surface assisted multiuser {MISO} systems,''
  \emph{IEEE Transactions on Vehicular Technology}, vol.~72, no.~6, pp.
  7653--7668, 2023.

\end{thebibliography}

\end{document}